\documentclass{jfm_arxiv}

\usepackage{natbib}
\usepackage{siunitx}
\usepackage{graphicx}
\usepackage{amsmath}
\usepackage{cleveref}
\usepackage{xspace}
\graphicspath{{../figurematter/latex/figs/}}

\newcommand{\kindex}[2]{\ensuremath{{#1}_{\scalebox{0.65}{#2}}}}
\newcommand{\Uinf}{\kindex{U}{$\infty$}\xspace}
\newcommand{\dd}[1]{\textrm{d}\ensuremath{#1}\xspace}

\newcommand{\numrangeand}[2]{\numrange[range-phrase = {\text{~and~}}]{#1}{#2}}

\newcommand\norm[1]{\left\lVert#1\right\rVert}

\title{On the wake and flapping dynamics of different aspect ratio flags}

\author{Gaétan Raynaud \and Karen Mulleners \corresp{\email{karen.mulleners@epfl.ch}}}

\affiliation{Unsteady Flow Diagnostics Laboratory, Institute of Mechanical Engineering, École Polytechnique Fédérale de Lausanne (EPFL), Lausanne, 1015, Switzerland}

\begin{document}
\maketitle
	
\begin{abstract}
The flapping of flags is a classical problem involving fast, large amplitude deformations of a thin flexible plate and unsteady flow phenomena. We perform systematic time- and space-resolved measurements of the deformation and drag acting on flapping flags for various aspect ratios and mass ratios. Flag deformations travel from the root to the tip while growing in amplitude. We present a robust method to determine the travelling wave properties, and reveal that the dimensionless wavelength $\lambda/L$ is constant for a given flutter mode, and that the wave speed equals the freestream velocity for aspect ratios above unity but decreases with decreasing mass and aspect ratio. We provide a mechanistic explanation of the influence of mass and aspect ratio on the flapping dynamics by introducing a mass-ratio-corrected dimensionless wave speed, which is proportional to the square root of the dynamic pressure difference across the flag. At smaller aspect ratio, the local dynamic pressure exerted by the fluid on the flag is reduced, leading to a decrease in the wave propagation speed and the tip frequency. Based on near wake flow field measurements, we identified two characteristic length scales, the ratio of flag area over its perimeter $L^*$ and the square root of area $\sqrt{HL}$ that scale the circulation of vortices shed per cycle. Values of the mean drag coefficient scatter widely, from \num{0} to \num{0.55}, when varying mass and aspect ratio. We can collapse the data when using a kinematics-based model with the mass ratio and the tip speed as inputs.
\end{abstract}

\section{Introduction}
\label{sec:introduction}

Flags are pieces of soft material used to display symbols, signals, or insignia and their typical waving in the wind makes them recognisable (Encyclopaedia Britannica, \citeyear{various_encyclopaedia_1910}).
The flapping flag problem and the various applications of flutter, the dynamic instability causing the motion of flags, have attracted research interest over many years.
Comprehensive reviews of the key contributions in this field are provided by \citet{shelley_flapping_2011} and \citet{yu_review_2019}. 
Flutter designates a dynamic fluid-structure instability where small deformations of an object in a fluid flow grow in amplitude to a limit cycle oscillation with constant frequency \citep{paidoussis_fluid-structure_2014}. 
The onset of flutter occurs when the flow speed exceeds a critical flow velocity such that the restoring effects of bending stiffness and flow-induced tension can no longer balance the inertial response of the flag, and the flag extracts energy from the flow \citep{connell_flapping_2007,shelley_flapping_2011}. 

The stability and the dynamics of flapping flags are governed by the interplay between inertia, bending rigidity, and fluid dynamic forces. 
The relative strengths of these contributing factors are described by several non-dimensional parameters.
The main dimensionless parameters that are relevant here are the reduced velocity, the mass ratio, the aspect ratio, and the Reynolds number.
Following the notations of \cite{eloy_flutter_2007}, these parameters are defined as:
\begin{equation}
U^* = \Uinf L \sqrt{\frac{\kindex{\rho}{s} e}{D}}, ~ M^* = \frac{\kindex{\rho}{f} L}{\kindex{\rho}{s} e}, ~\frac{H}{L}, \text{  and  } Re = \frac{\Uinf L}{\nu},
\end{equation}
for a flag of length $L$, height $H$, thickness $e$, density $\kindex{\rho}{s}$, and bending rigidity $D$ in an incoming fluid flow of velocity $\Uinf$, fluid density $\kindex{\rho}{f}$, and kinematic viscosity $\nu$.

The reduced velocity $U^*$ compares the characteristic period of free vibrations of a beam in vacuum $\sim L^2 \sqrt{\kindex{\rho}{s}e/D}$ to the convective time of the surrounding flow $L/\Uinf$ \citep{virot_fluttering_2013}.
The mass ratio $M^*$ compares the magnitude of the fluid mass displaced during a flapping cycle of the flag $\sim \kindex{\rho}{f} HL^2$ to the mass of the flag itself $\kindex{\rho}{s} e L H$.
The mass ratio can also be seen as the prefactor of the normalised pressure forces term in the spanwise-averaged momentum balance \citep{eloy_flutter_2007}.
Large mass ratios correspond to flags in a dense fluid, or light flags.
The aspect ratio $H/L$ compares the height to the length of a flag.
Flags in the limit of low aspect ratio $H/L \rightarrow 0$ behave like ribbons \citep{lemaitre_instability_2005}.
Two-dimensional conditions are approximated by large aspect ratio flags $H/L \gg 1$ and have been a classical simplification in the study of flag stability \citep{alben_dynamics_2022,connell_flapping_2007,michelin_vortex_2008}.
The Reynolds number indicates the relative importance of inertial versus viscous effects in the flow, and the influence of the Reynolds number on the flapping dynamics of flags is expected to be negligible for $Re > 1000$ \citep{yu_review_2019}.

The stability boundary of the flag is marked by the critical reduced velocity. 
Analytical models of flags have focused on predicting the critical reduced velocity as a function of mass ratio and aspect ratio.
The critical velocity is most sensitive to the mass ratio and the value of the mass ratio governs which flapping mode becomes unstable at the flapping onset \citep{yu_review_2019}.
At low mass ratio ($M^*< \mathcal{O}(1)$), the critical reduced velocity $U^*$ is a decreasing function of $M^*$.
For higher mass ratios ($M^*>1$), the critical reduced velocity settles around $U^* \sim \num{10}$ and different branches describe the evolution of $U^*$ with $M^*$ depending on the unstable flapping mode that is excited \citep{michelin_vortex_2008, eloy_aeroelastic_2008}.
The minimum values of the critical velocity for different branches are found for infinite aspect ratio flags.
Finite aspect ratio flags are subject to three-dimensional flow effects that reduce the pressure forces and delay the onset of flapping to higher reduced velocities.
The critical reduced velocity tends to increase with decreasing aspect ratio $H/L$ \citep{eloy_flutter_2007, kumar_effect_2024,chun2010flutter}.
Overall, analytical models tend to underestimate the critical velocity compared to experimental results \citep{yu_review_2019}, and critical velocities obtained from experiments show some inherent fluctuations, some of which are due to flatness defects \citep{eloy_origin_2012}.

Flags maintain large amplitude flapping for reduced velocities beyond the onset velocity, which is referred to as the post-critical regime \citep{tang_instability_2007}.
The spatial characteristics of post-critical flapping are extracted from the flapping amplitude envelope.
The first person to experimentally study the amplitude envelope through stroboscopic imaging and to identify a travelling deformation wave propagating towards the tip was \cite{taneda_waving_1968}. 
Depending on mass ratio, different flapping envelopes are observed for different flapping modes \citep{taneda_waving_1968}. 
Necks are used to describe the spatial evolution of the flapping envelope.
Necks are regions of the flag where the envelope has locally a smaller amplitude.
The number of necks along the flag increases with higher order flapping modes and changes the bending distribution along the flag.
An accurate localisation of the high bending regions along the flag is important to design energy harvesters based on piezoelectric patches \citep{doare_piezoelectric_2011}, where the typical milliwatt output power is relevant to supply wireless sensor networks \citep{yu_energy_2016} and for battery-less flow sensors \citep{liu_development_2012}. 

The main temporal characteristic of post-critical flapping is the flapping frequency.
The tip frequency $f$ increases linearly with $\Uinf/L$, which is the inverse of the convective time \citep{yu_review_2019}.
The ratio $fL/\Uinf$ corresponds to the Strouhal number and typically ranges between \numrangeand{0.1}{1} \citep{taneda_waving_1968}.
For a given flag geometry, the Strouhal number remains almost constant when varying the wind velocity unless the flutter mode changes \citep{virot_fluttering_2013}.
The tip frequency and Strouhal number $fL/\Uinf$ decrease, for decreasing aspect ratios at constant mass ratios \citep{huang_three-dimensional_2010,kumar_effect_2024}.
The decrease in flapping frequency with aspect ratio is linked to a drop in the pressure difference for lower aspect ratio flags due to the formation of flow structures at the edges, which reduce the driving force of the flapping \citep{huang_three-dimensional_2010}. 
More details on the origin of the flapping time scales are desirable for further investigation of various applications involving membrane flutter, including, for instance, the diagnostics of respiratory wheezes \citep{gavriely_flutter_1989}, orinasal snoring and sleep apnoea syndromes \citep{huang_flutter_1995,auregan_snoring_1995} where the frequency of pressure waves is perceived as an audible sound.
Understanding the effect of aspect ratio and mass ratio on the flapping timescales also matters because human tissues come in a variety of shapes and thicknesses.

During the large amplitude flapping, vorticity is accumulated along the surface on both sides of the flags and released alternatingly during the upstroke and downstroke motion of the tip \citep{kumar_dynamics_2021}.
Three-dimensional flow structures are characterised by an alternating release of horse-shoe shaped vortex tubes from the trailing edge, and the wake structures weaken with decreasing aspect ratio \citep{huang_three-dimensional_2010,kumar_effect_2024}.
The time-varying wake produced during flutter serves practical applications such as micro-scale mixers on lab-on-chip devices \citep{rips_enhanced_2019}, or low-Reynolds number heat dissipation systems \citep{shoele_computational_2014}.
The wake past a fluttering leaflet of an aortic valve prosthesis generates higher levels of shear stress compared to a non-fluttering leaflet, contributing to the formation of thrombus \citep{bornemann_leaflet_2025}.
Functional relationships between the shape properties and wake quantities would help design biomedical implants and mixing devices matching the desired flow properties.

A sudden increase in drag force is observed as the flag starts flapping \citep{taneda_waving_1968}.
The origin of the drag increase has been associated with the dynamically-induced tension of the oscillating flag \citep{moretti_tension_2003}, and with flow reorientation in the alternating wake \citep{muller_fish_2003}.
Strong variations are observed in the mean and fluctuating drag coefficients between different flags and during different flapping regimes.
\cite{taneda_waving_1968} reports drag coefficient values that range from \num{0.02} up to \num{2}.
The efforts to model the drag have focused on quantities derived from the tip amplitude and frequency \citep{virot_fluttering_2013}.
The increase in the forces experienced by a flapping flag compared to a non-flapping flag are responsible for material damage, and can deteriorate the production quality in fast-operating industrial printing \citep{watanabe_experimental_2002}.
The mean drag coefficient of different flags can vary by two orders of magnitude, and more than half a century after the seminal work from \cite{taneda_waving_1968}, we still do not have sufficient intuition or a physical explanation for why this is the case \citep{vogel2009leaves}.

In this paper, we want to change this and provide an overall better understanding of the interaction between flag deformation, wake parameters, and mean force. 
In particular, we focus on the effect of aspect ratio on the post-critical flapping regime of rectangular flags.
We first review the flag instability and describe the spatial characteristics of the double-neck flutter regime.
We present a robust method to quantify the propagation rate of the travelling wave and compute an effective wave length.
Variations in flapping timescales, tip speed, and wave propagation speeds are investigated through systematic measurements of deformation after the onset of flapping.
The analysis of the wave speed is the starting point to understand the range of measured flapping time scales, and we investigate physical models that take aspect ratio effects into account.
Flow measurements of the close wake show how these changes in timescales and speeds affect the formation of the vortex wake.
Mean drag data is measured and presented for various reduced velocities.
We discuss and test a predictive model for the mean drag coefficient based on the tip velocity.
Thanks to the combined deformation, flow field, and drag measurements, we will provide physical insights into why different flags can have such vastly different mean drag coefficients.

\section{Experimental methods}

\subsection{Rectangular flags}

Rectangular flags were made out of white paper (PAPETERIA A4 $\qty{80}{\gram\per\meter\squared}$).
Flags were designed with alignment holes to be clamped on a flag pole and held with the long paper side along the stream-wise direction (\cref{fig:setup}a).
The flag pole clamps the paper flag between two blades of stainless steel with an isosceles trapezoidal cross-sections.
The blades have a maximum thickness of \qty{5}{\milli\meter} and a streamwise length of \qty{30}{\milli\meter}.
The leading and trailing edges are sharpened by \ang{30} bevel edges of \qty{8.75}{\milli\meter} long.
The rigidity of the A4 paper sheets in the long direction was measured using an elasto-gravity bending procedure, resulting in $D = Ee^3/12(1-\kindex{\nu}{s}^2) = \qty[separate-uncertainty]{3.4(0.4)e-4}{\newton\meter}$, where $E$ and $\kindex{\nu}{s}$ are the Young modulus and the Poisson ratio. 
The uncertainty of $\qty{0.4e-4}{\newton\meter}$ corresponds to the standard variation between different samples, and the individual measurement error of the procedure is $\qty{0.06e-4}{\newton\meter}$.
A total of \num{48} shapes were cut using a desktop laser cutter with various heights $H$ ranging from \qtyrange{4}{19.6}{\centi\meter} and lengths $L$ ranging from \qtyrange{10}{20}{\centi\meter}  (\cref{fig:setup}b).
As the flag length is present in the definition of the mass ratio and the aspect ratio, we tested different flag heights for each flag length such that we have several flags with the same aspect ratio, but different mass ratio and vice versa.
All tested flags fall in the aspect ratio range $\num{0.22}<H/L<\num{1.92}$ and mass ratio range $1.4<M^*=\kindex{\rho}{f}L/\left(\kindex{\rho}{s}e\right) <2.8$.
They are tested for reduced velocities $7<U^* = \Uinf L \sqrt{\kindex{\rho}{s}e/D}<31$.
For all the experiments, the Reynolds number $Re = \Uinf L / \nu$ falls between \numrange[range-phrase = {\text{~and~}}]{3e4}{1.5e5}. 

The room conditions for every measurement at all flow velocities were systematically monitored using an integrated temperature, ambient pressure, and humidity sensor (BME688) that was placed right outside the test section.
The average room temperature was \qty{20.8}{\degreeCelsius}, with a standard deviation of \qty{0.25}{\degreeCelsius}, below the rated accuracy of the sensor (\qty{0.5}{\degreeCelsius} in the range \qtyrange{0}{65}{\degreeCelsius}).
The minimum measured temperature was \qty{19.8}{\degreeCelsius}, and the maximum was \qty{21.3}{\degreeCelsius}.
The room humidity was comprised between \qty{25}{\percent} and \qty{37}{\percent}.
Based on the measured overall small variations of temperature and low values of room humidity, we expect that mechanical properties of the paper are not significantly affected by the room conditions. 

\subsection{Experimental set-up}

\begin{figure}
\includegraphics[width=\linewidth]{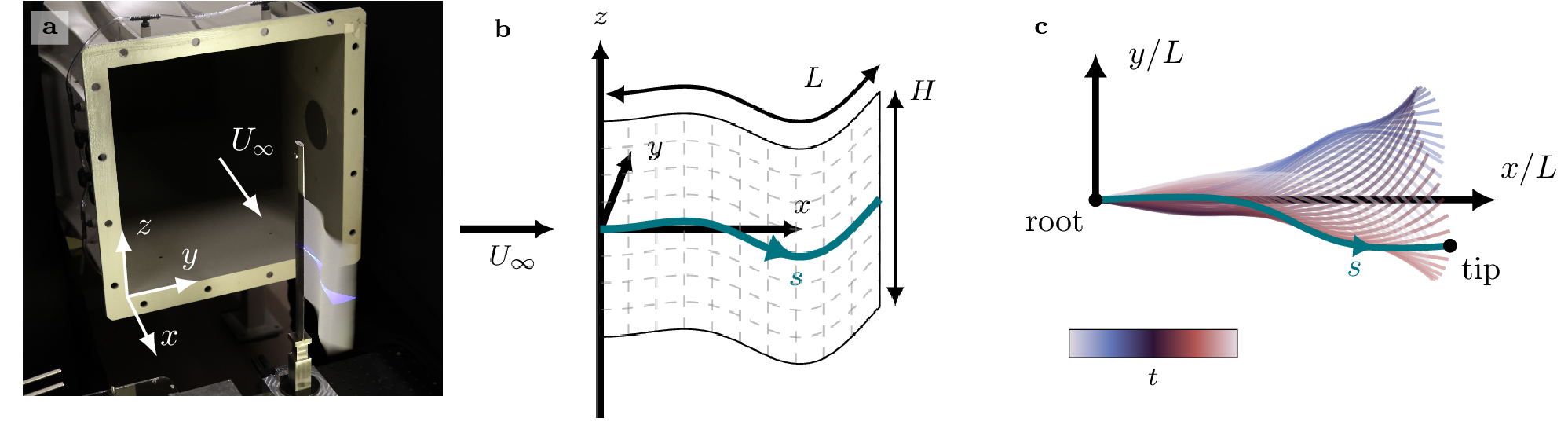}
\caption{(a) Photograph of the experimental setup.
(b) Sketch of the flag with its centreline highlighted with $s=0$ at the root and $s=L$ at the tip. 
(c) Exemplary visualisation of the time-resolved measurements of the deformation of the centreline.}
\label{fig:setup}	
\end{figure}

The experiments are conducted in a wind tunnel with an open section of \qtyproduct{45 x 45}{cm}.
Flags are held vertically at the centre of the flow stream (\cref{fig:setup}a).
For each flag, a measurement series consists of gradual steps of flow velocity $\Delta \Uinf$.
The flow velocity is first increased from rest to the onset of flapping with $\Delta \Uinf = \qty{0.2}{\meter\per\second}$.
Once flapping is reached, the flow velocity is decreased to capture the hysteresis with smaller steps of $\Delta \Uinf = -\qty{0.1}{\meter\per\second}$.
The velocity at which the flag restabilises defines the offset velocity. 
Then, the velocity is increased again to above the onset velocity to gather additional data in the post-critical regime with velocity increases of $\Delta \Uinf = \qty{0.2}{\meter\per\second}$. 

At every flow velocity, the spatiotemporal deformation of the centreline is measured (\cref{fig:setup}b).
A continuous laser pointer and a Powell lens generate a horizontal light sheet at the flag mid-height.
The reflection of the centreline of the flag is recorded with an event-based camera (Century Arks SilkyEvCam VGA) for a duration of $\qty{6}{s}$.
The planar deformations are reconstructed from the raw stream of events using an iterative spatial-clustering algorithm \citep{raynaud_event-based_2025}.
The reconstruction error for this methodology is of the order or below one percent of the flag length.
As output of the deformation reconstructions we obtain series of $x$ and $y$ coordinates sampled at \num{50} points uniformly distributed along the flag centreline (\cref{fig:setup}c).
The reconstruction update rate of the output deformation is set during post-processing with a range of values between \qtyrange{1.5}{2}{\kilo\hertz} depending on the rate of events. 
The flapping frequency is determined by applying a fast Fourier transform (FFT) of the tip deformation signal. 
The uncertainty associated with the flapping frequency is the sum of the frequency resolution of the FFT and the standard deviation of the peak frequency when the FFT is applied to different sub-intervals of the recording.

The flag pole is attached to a force platform that uses three single-axis beam load cells.
The load cells are composed of strain gauges in a Wheatstone bridge formation, connected to a custom power supply and voltage amplification circuit.
An analogue to digital converter (National Instruments NI-9215) records force data for a duration of \qty{10}{\second}, at a sampling rate of $\kindex{F}{s} = \qty{2}{\kilo\hertz}$.
The load cell is manually calibrated over a range of $\pm \qty{2}{\newton}$.
The discretisation error from the \num{16} bit digital converter is lower than \qty{0.7}{\milli\newton} and the mean residual error is \qty{10}{\milli\newton} after calibration.
The total uncertainty for the mean drag force and mean drag coefficient is the sum of the discretisation error, the calibration uncertainty, and the repeatability of the mean drag value computed on independent segments of the recording, and has an average value of \qty{16}{\milli\newton} for our dataset where the mean drag ranges from \qtyrange{0}{1.13}{\newton}. 
The mean drag on the pole without a flag is recorded at different flow velocities, and subtracted from the mean drag measured on the pole with a flag such that \kindex{F}{x} in \cref{subsec:dragscaling} refers to the fluid force acting on the flag only.

\subsection{Particle image velocimetry}

Flow data are gathered for a selection of flags of length $L=\qty{16}{cm}$ ($M^*=\num{2.27}$) and aspect ratios ranging from \numrange{0.375}{1.25}.
A low repetition rate laser (Quantel Evergreen 200) illuminates the mid-height plane in the region around the tip and the close wake of the flag.
The wind tunnel room is seeded using a fog machine (Hazebase highpower) and a glycol-water mixture.
A camera (PCO Edge 5.5) with a lens (Canon EF \qty{50}{\milli\meter} f/1.4) is positioned above the flag and records series of double images.
The optical axis of the camera is located downstream of the flag such that flow close to the tip is observed without optical obstruction by the flag itself.
A synchroniser (ILA Synchronizer 2011) triggers the camera and the laser.
The synchroniser controls both the inter-frame timing and the power settings of the laser.
A total of \num{2000} image pairs are gathered at a flow speed of $\Uinf=\qty{7.5}{\meter\per\second}$, which corresponds to $U^*=\num{18.3}$ when the length is fixed to $L=\qty{16}{\centi\meter}$.
We set the inter-frame duration to $dt=\qty{55}{\micro\second}$ and the pulse energy to \qty{60}{\milli\joule} for all the cases. 
A standard multi-grid algorithm processes the particle images \citep{raffel_particle_2007}.
The final window size of \qtyproduct{48x48}{px} and an overlap of \qty{50}{\percent} results in a spatial discretisation of $\eta = \qty{2.7}{mm}$ or $\eta/L = \num{1.7e-2}$.
Individual snapshots are phase-averaged.
The phase of the flapping is determined using the corresponding event-based deformation measurements.

\section{Results and discussion}
In this paper, we want to provide insight into the effect of aspect ratio on the interaction between flag deformation, wake parameters, and the mean force of flapping flags. 
In particular, we focus on the post-critical flapping regime which corresponds to the range of flow velocities beyond the critical velocity. 
First, we revise how the aspect ratio affects the critical velocity (\cref{subsec:instabilitymap}).
Then, we focus on the spatial properties of the flapping kinematics (\cref{subsec:lengthscales}) before investigating the temporal properties and their variations with aspect ratio and mass ratio (\cref{subsec:timescales}).
We aim to provide a physical explanation of the variation of flapping timescales reported in the literature.
To support this investigation, we introduce a robust method to quantify and characterise the spatiotemporal properties of the flag deformation waves.
We further report variations of wake quantities and propose scaling lengths that incorporate the three-dimensional effects (\cref{subsec:wakequantities}).
Finally, we focus on the force data to understand how drag scales with flow velocity depending on the flapping regime and the flag properties (\cref{subsec:dragscaling}).
To provide a better understanding for the broad range of mean drag coefficients observed, we combine the flapping time scales results into an empirical model that quantitatively describes and explains the spread of the drag data across flapping regimes, mass ratio, and aspect ratio.

\subsection{Critical flapping velocity}
\label{subsec:instabilitymap}

\begin{figure}
\includegraphics[width=\linewidth]{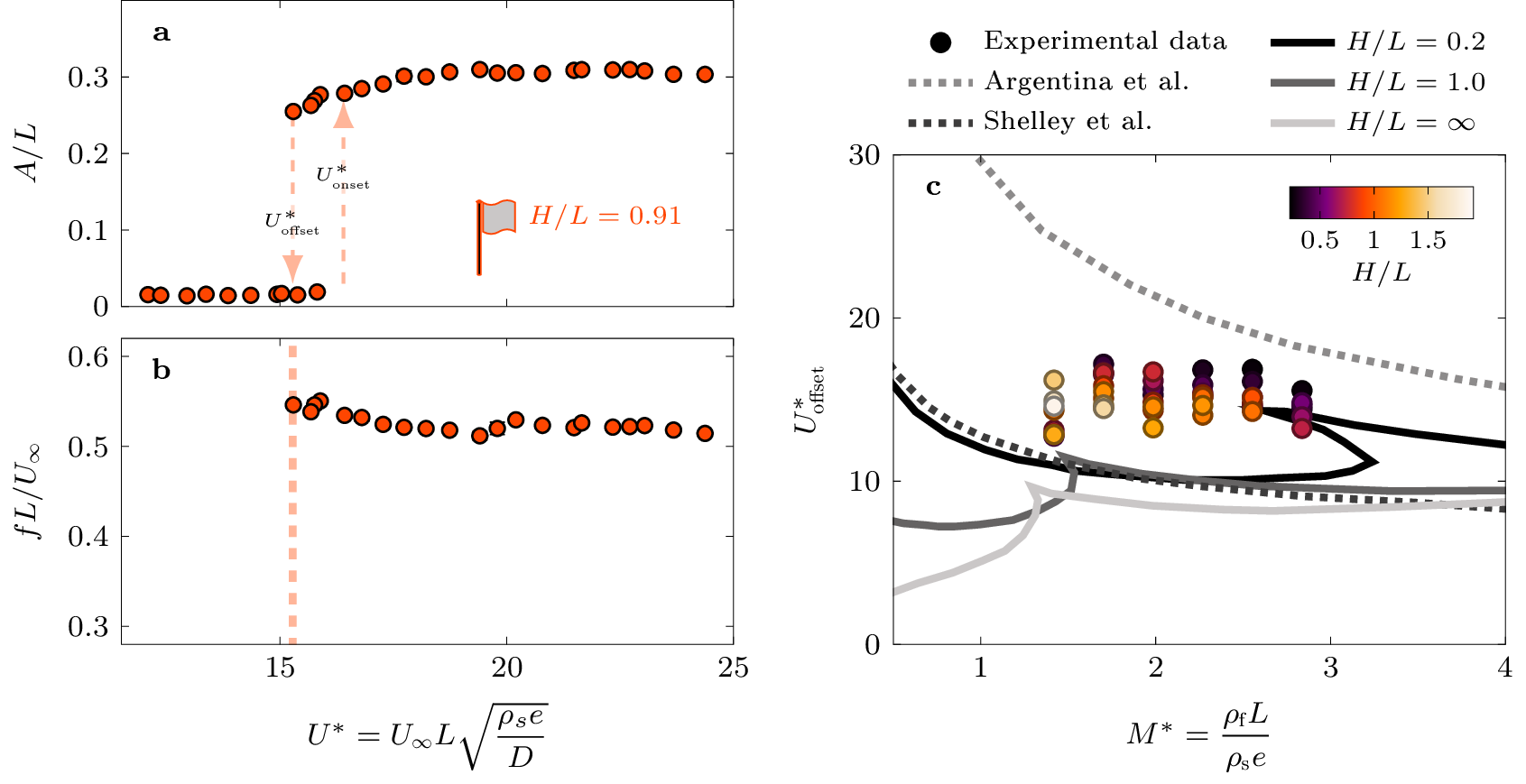}
\caption{(a) Dimensionless tip amplitude $A/L$ and (b) dimensionless flapping frequency $fL/\Uinf$ as a function of reduced velocity $U^*$ for a flag of length $L=\qty{16}{cm}$ and height $H=\qty{14.5}{cm}$ corresponding to $H/L = \num{0.91}$. 
The vertical dashed line in (b) represents the reduced offset velocity. 
(c) Reduced offset velocities $\kindex{U}{offset}^*$ for the 48 tested flags compared with instability thresholds from the analytical models of \cite{argentina_fluid-flow-induced_2005, shelley_heavy_2005} for infinite aspect ratio flags and from \cite{eloy_flutter_2007} for flags with three aspect ratios.}
\label{fig:ClassicFlagQuantities}	
\end{figure}

Fluttering flags undergo periodic large amplitude deformations that can be characterised by a dimensionless tip amplitude and frequency.
The tip amplitude $A$ is measured as the maximum transverse deflection of the tip, $\text{max}\left|\kindex{y}{tip}\right|$ with respect to the midline $y=0$.
The flapping frequency $f$ is obtained from the peak of the fast Fourier transform of the tip displacement measurement signal.
In \cref{fig:ClassicFlagQuantities}a and b, we show the dimensionless tip amplitude and frequency as a function of the reduced velocity for an example flag of $M^*=\num{2.27}$ and $H/L=\num{0.91}$.
At low velocities ($U^*<\num{15}$), the flag stands still, the tip amplitude is almost \num{0} (\cref{fig:ClassicFlagQuantities}a) and the flapping frequency is not defined (\cref{fig:ClassicFlagQuantities}b).
The dimensionless tip amplitude $A/L$ suddenly increases from almost \num{0} to \num{0.27} when we increase the reduced flow velocity beyond the critical onset velocity $\kindex{U}{onset}^* = \num{16.4}$ (upward arrow in \cref{fig:ClassicFlagQuantities}a).
The amplitude plateaus at $A/L = \num[separate-uncertainty]{0.31(0.01)}$ when further increasing $U^*>18$.
In this post-critical region, the dimensionless tip frequency is also approximately constant $fL/\Uinf = \num[separate-uncertainty]{0.53(0.01)}$ when varying the reduced velocity $U^*$.
When the velocity is reduced starting from the post-critical regime, the flag restabilises.
The velocity at which the flag restabilises is called the offset velocity and is typically smaller than the onset velocity.
Initial sample curvature and flatness defects can lead to delayed onset of flutter but have only a minimal influence on the critical offset velocity \citep{eloy_origin_2012,mettelsiefen.2025}. 
The extraction of the flapping offset is more repeatable, and we preferably use the critical offset velocity for comparisons with experimental and theoretical results from literature.

For the example presented in \cref{fig:ClassicFlagQuantities}a, there is a clear jump in the amplitude when we drop below the critical offset velocity. 
For other flags in our data set, we observe a more gradual decrease in the tip amplitude when we approach the offset velocity, which is more similar to what is observed by \cite{eloy_origin_2012}.
The evolution of the flapping amplitude with flow velocity close to the flapping offset varies with the aspect ratio and the mass ratio.
\cite{hiroaki2024flutter} observed that the jump in amplitude reduces with decreasing aspect ratio. 
A detailed analysis of the evolution of the tip amplitude near the critical velocity would be interesting but exceeds the scope of this work. 
We extract and briefly discuss the critical velocities of our flags in the next paragraph before shifting our focus to the post-critical flapping dynamics. 

We systematically extracted the critical reduced offset velocity for all tested flags as the lowest flow velocity with a significant flapping amplitude $A/L>0.1$.
A sensitivity analysis done on this amplitude threshold reveals a minor influence on the extracted critical velocities for  threshold values between $A/L=0.05$ and $A/L<0.15$. 
The reduced offset velocity $\kindex{U}{offset}^*$ lies between \num{12.78} and \num{17.17} for different mass ratios $\num{1.4} < M^* < \num{2.8}$ and aspect ratios $\num{0.22}<H/L<\num{1.92}$ (\cref{fig:ClassicFlagQuantities}c).
The experimentally extracted offset velocities are compared with three analytical models in the $(M^*,U^*)$ map.
Linear models find the smallest flow velocity for which the flag deformations are not damped, so that vibration can sustain, which describes the transition at the offset velocity.
The three analytical models predict only the critical offset velocity $\kindex{U^*}{offset}$ \citep{eloy_origin_2012}.
Two linear two-dimensional potential flow models from \cite{argentina_fluid-flow-induced_2005} and \cite{shelley_heavy_2005} predict that the critical reduced velocity decreases monotonically with increasing mass ratio.
Our experimental data points lie between the two linear models and do not vary significantly with mass ratio.
A three-dimensional potential flow model from \cite{eloy_flutter_2007} takes into account the effect of aspect ratio and predicts several modal branches for different values of mass ratio.
In the range of $1.4<M^*<2.8$, which corresponds to our experimental data range, the first unstable mode is the double-neck flutter \citep{eloy_aeroelastic_2008}.
In the region of the double-neck flutter, the predicted reduced offset velocities for three aspect ratios $H/L = \num{0.2}$, \num{1}, and $+\infty$ lie between $8<\kindex{U}{offset}^*<11.5$ and slightly decrease with aspect ratio.
Our experimental values are almost \num{1.5} times larger than the predicted values of \cite{eloy_flutter_2007} for the  flutter mode. 
We do not observe a clear variation with aspect ratio.
The experimentally determined values of the critical reduced velocity are often larger than the theoretical predictions and show inherent fluctuations \citep{connell_flapping_2007,jankee_influence_2022,yu_review_2019}.
The onset velocities are sensitive to initial sample curvature, flatness defects, and other sample imperfections.
The imperfections increase the cross-sectional stiffness of the flags when they are not flapping and increase the measured onset velocities \citep{eloy_origin_2012,mettelsiefen.2025}. 
The reduced critical offset velocities from our experiment and from the three-dimensional model are driven by the two neck flutter regime and only weakly affected by aspect ratio and mass ratio.

\subsection{Spatial characteristics of the deformation waves}
\label{subsec:lengthscales}

\begin{figure}
\includegraphics[width=\linewidth]{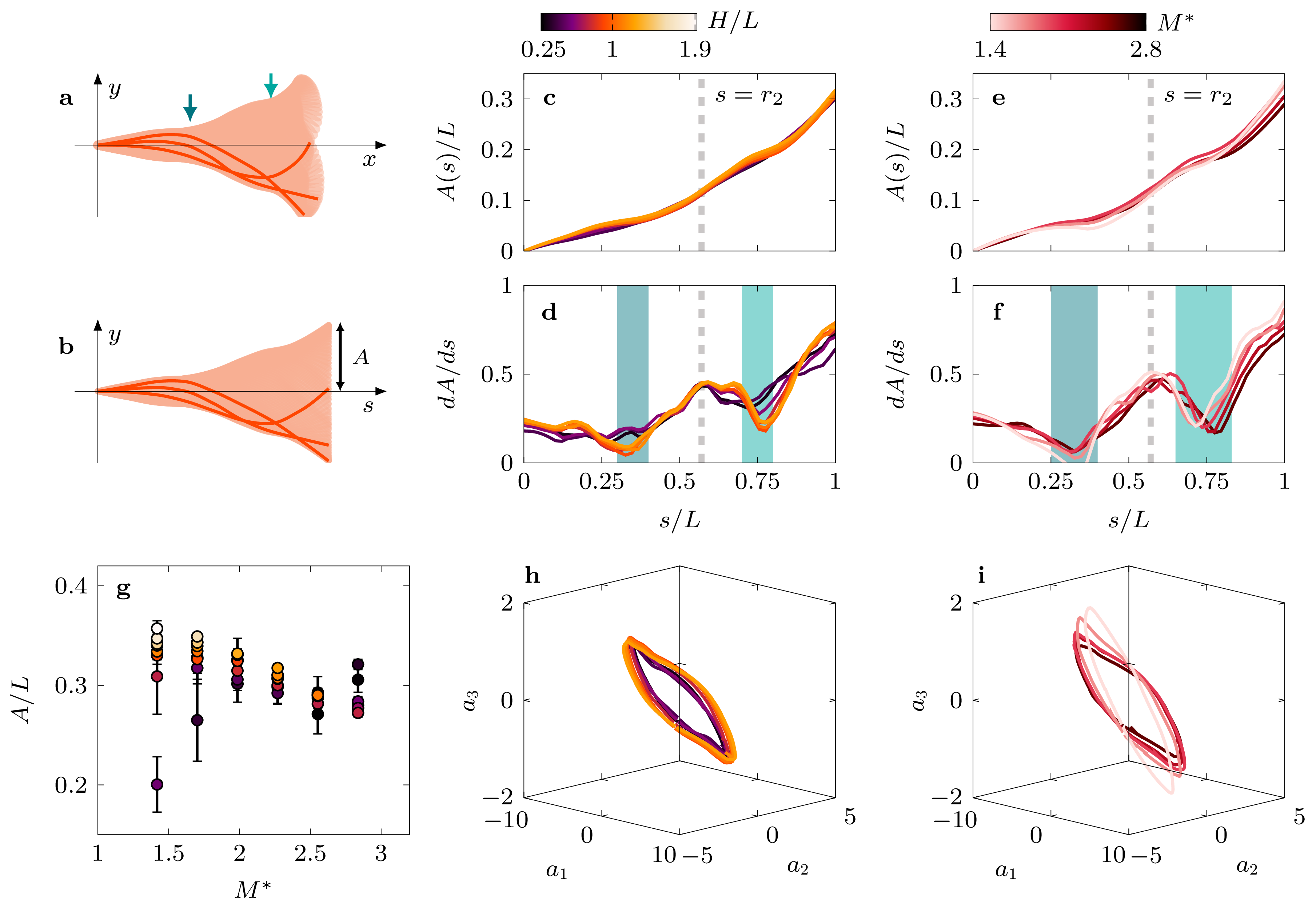}
\caption{(a) Superposition of the time-resolved centreline deformations for a flag of $M^* = \num{2.27}$ and $H/L = 0.91$ with individual deformations highlighted by the darker lines. 
The two arrows indicate the position of the two necks for this double flutter mode. 
(b) Superposition of the time-resolved transverse deformation along the curvilinear abscissa $s$. 
(c, e) Envelope of the flapping amplitude in dimensionless coordinates $A(s)/L$ as a function of $s$. 
(d, f) Spatial gradient of the envelope $dA/ds$. 
The shaded areas indicate the curvilinear position range where dips in the envelope gradient occur. 
(g) Dimensionless tip amplitude $A/L$ for each flag plotted against mass ratio and coloured by aspect ratio. 
For each flag, the amplitude is the average value of the measurements at different post-critical velocities and the error bar indicates the standard deviation. 
(h, i) Phase-averaged evolution of the time coefficients $(\kindex{a}{1},\kindex{a}{2},\kindex{a}{3})$ of the three first beam modes. 
Flags of same mass ratio $M^* = \num{2.27}$ and different aspect ratio are compared in (c, d, and h).
Flags of similar aspect ratio $H/L \approx 1$ and different mass ratio are compared in (e, f, and i).}
\label{fig:EnvelopeAndBeamModes}	
\end{figure}

In the post-critical regime, the dimensionless tip-based quantities are approximately constant (\cref{fig:ClassicFlagQuantities}a,b).
At a given post-critical reduced velocity, the flag periodically moves from one side to the other, and the transverse motion propagates from the root to the tip. 
First, we investigate the envelope of the motion in \cref{fig:EnvelopeAndBeamModes}a-f. 
In \cref{fig:EnvelopeAndBeamModes}a, the centreline deformations in the Cartesian coordinates $(x,y)$ are stacked on top of each other to create the envelope of the motion for one example flag ($H/L = 0.91$ at $M^* = \num{2.27}$).
In \cref{fig:EnvelopeAndBeamModes}b, we now stack the transverse deformation as a function of the curvilinear coordinate $s$ for the same example flag at the same conditions.
The local amplitude is zero at the root, and increases from the root to the tip.
Two regions of locally smaller lateral extent are highlighted with arrows and correspond to the two necks.
To extract the position of the necks, we use the spatial evolution of the local amplitude along the centreline.
The amplitude along the centreline is computed as the maximum absolute transverse displacement $A(s) = \max_{t} \left|y(s,t)\right|$.
In \cref{fig:EnvelopeAndBeamModes}c, we show the local amplitude along the centreline for different aspect ratios at $M^*=\num{2.27}$.
The local amplitude increases monotonically along the curvilinear coordinate for all tested aspect ratios.
Three regions of almost constant amplitude growth are separated by two dips in the $dA/ds$ plot, which correspond to the positions of the necks at $s/L \approx 0.35$ and $s/L \approx 0.75$ (\cref{fig:EnvelopeAndBeamModes}d).
Between the two necks, a local maximum of the amplitude growth $dA/ds$ occurs.
The location of this local maximum coincides with the radius of the second moment of area $s=\kindex{r}{2} = L/\sqrt{3}$.
In \cref{fig:EnvelopeAndBeamModes}e we show the local amplitude along the centreline for flags of different mass ratios and $H/L\approx \num{1}$.
The local amplitude also increases along the centreline for all tested mass ratios, but now we observe larger differences between flags than when we compared flags of different aspect ratios in \cref{fig:EnvelopeAndBeamModes}c,d.
When the mass ratio is constant, the location of the necks is approximately constant over the range of aspect ratios tested.
The two necks move slightly closer to the tip with increasing mass ratio (\cref{fig:EnvelopeAndBeamModes}f).
At the tip ($s=L$), the maximum dimensionless amplitude $A/L$ and the amplitude slope $dA/ds$ decrease with increasing mass ratio (lighter flags, \cref{fig:EnvelopeAndBeamModes}e,f) and they are constant for different aspect ratios (\cref{fig:EnvelopeAndBeamModes}c,d).
The tip amplitudes of the post-critical limit cycle oscillations are averaged over different reduced velocities.
These averaged tip amplitudes of all tested flags are presented as a function of mass ratio in \cref{fig:EnvelopeAndBeamModes}g.
For a given mass ratio, the tip amplitude is approximately constant across all tested aspect ratios, except for two low aspect ratio flags at $M^*=\num{1.4}$ and \num{1.7}.
When the mass ratio is increased from $M^*=\num{1.4}$ to $M^*=\num{2.8}$, which corresponds to going from heavier to lighter flags, the average tip amplitude decreases from $A/L = 0.35$ to $A/L = 0.28$.
Overall, for all flags tested here, the envelope data show the same double-neck flutter mode in the post-critical flapping regime.
The aspect-ratio does not seem to have a significant influence on the spatial characteristics of the flapping deformation for the range of mass ratio tested here.  
The mass ratio affects the position of the necks and the tip magnitude.
With increasing mass ratio, i.e. considering lighter flags, the necks move towards the tip and the tip amplitude decreases. 

To further highlight the influence of mass ratio and aspect ratio on the spatial flapping deformations, we create a reduced order representation of the flapping cycle based on the three first cantilevered beam mode shapes.
We obtain time-varying mode coefficients describing the flapping cycle by projecting the measured time-resolved transverse displacements $y(s,t)$ along the centreline on the beam mode shapes $\kindex{\psi}{i}$: 
\begin{equation}
\kindex{a}{i} (t) = \frac{1}{L^2}\int\limits_{s=0}^L \kindex{\psi}{i}(s) y(s,t) \dd s, ~ i = 1,2,3\quad.
\end{equation}
In \cref{fig:EnvelopeAndBeamModes}h, we show the flapping cycle evolution of the phase-averaged coefficients $(\kindex{a}{1},\kindex{a}{2},\kindex{a}{3})$ for flags of different aspect ratios.
The typical cycle is a single loop in the cantilever mode space and all the trajectories are almost identical despite the variations in aspect ratio.
This confirms that aspect ratio does not affect the spacial characteristics of the flapping for the range of aspect ratios considered here.  
In \cref{fig:EnvelopeAndBeamModes}i, we show the phase-averaged coefficients for flags of different mass ratios.
The typical cycle is also a loop, but there are noticeable differences between the trajectories.
The coefficient of the third beam mode $\kindex{a}{3}$ reaches larger values when reducing the mass ratio, i.e. for heavier flags.
The three first beam mode shapes are included in the appendix (\cref{app:beammodes}, \cref{fig:appCantileveredModes}) together with two-dimensional projections of the coefficient loops.
The third mode shape has a stronger bending towards the tip compared to the first two mode shapes.
An increase in the contribution of the third beam mode indicates larger curvature along the centreline and explains the larger values of $dA/ds$ close to the tip when reducing mass ratio (\cref{fig:EnvelopeAndBeamModes}f).
The limit cycle oscillation of the flag deformation can be reconstructed as a combination of the cantilevered beam mode shapes. 
As the trajectory in the $(\kindex{a}{1},\kindex{a}{2},\kindex{a}{3})$ space is a single loop with intersection, the contributions of the different modes have the same oscillation frequency and a phase offset with respect to each other, characteristic of a wave travelling along the flag.

\begin{figure}
\includegraphics[width=\linewidth]{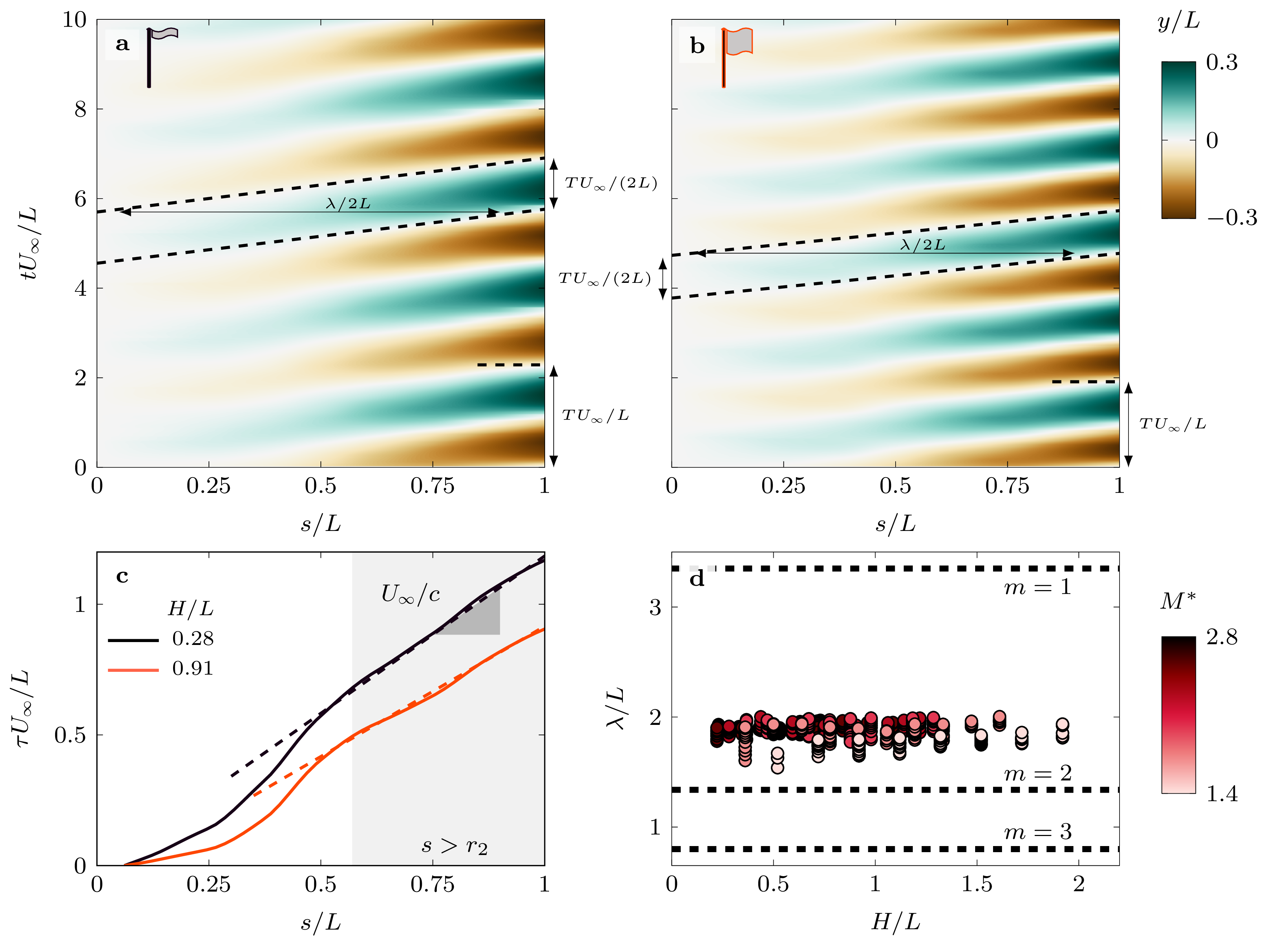}
\caption{(a, b) Space-time diagrams of the lateral displacement $y/L$ for 10 convective times for two flags (a) $H/L=\num{0.28}$ and (b) $H/L=\num{0.91}$, both at $M^*=\num{2.27}$. 
(c) Correlation delay along the curvilinear coordinate for the data shown in (a) and (b). 
The slope in the shaded area corresponding to $s>\kindex{r}{2}$ is extracted to get the corresponding wave speed $c$ and wavelength $\lambda = c/f$. 
(d) Wave length $\lambda/L$ as a function of aspect ratio for all tested flags. 
The dashed horizontal lines indicate the dimensionless wavelengths of the three first cantilevered beam modes.
}
\label{fig:SpatialDeformations}	
\end{figure}

The properties of the travelling wave are extracted based on the full spatiotemporal deformation data of the centreline.
We can visually represent the transverse deformation $y/L$ along the centreline as a function of the curvilinear coordinates $s/L$ and normalised convective time $t \Uinf/L$ in the form of space-time diagrams. 
The space-time diagrams for two example flags, $H/L=\num{0.28}$ and $H/L=\num{0.91}$, are represented in \cref{fig:SpatialDeformations}a and b, respectively. 
The mass ratio for both example flag is $M^*=\num{2.27}$.
Both space-time diagrams feature diagonal patterns with a positive slope, indicating the propagation of a travelling wave from root to tip.
The absolute value of the local displacement increases along the diagonal lines, indicating that the local amplitude grows as the wave propagates.
The vertical spacing between two consecutive $y=0$ contour lines corresponds to half of the dimensionless flapping period.
The horizontal spacing corresponds to half the dimensionless wavelength, i.e. $\lambda/2L$, and represents the distance along the flag between two opposite signed crests.
The slope of the diagonals corresponds to $\lambda f/\Uinf$, with  $\lambda f$ the wave speed which we denote by $c$. 

The wave length and wave speed have not received much attention to date, despite the crucial role they play in the mechanistic explanation of the flapping dynamics, as we will demonstrate in the next sections. 
To systematically and reliably extract the wave length and wave speed, we correlate the temporal evolution of the measured flag deflection at different locations along the centreline.
We introduce the correlation delay $\tau$ (\cref{fig:SpatialDeformations}c) as a measure of the propagation delay of the wave pattern along the flag.
The time delay $\tau$ at a given position $s$ with respect to a reference position $\kindex{s}{ref}$ is obtained using the maximum of the cross correlation of $y(s,t)$ and $\kindex{y}{ref}(t) = y(\kindex{s}{ref},t)$:
\begin{equation}
\tau(s) = \text{argmax}_{\tilde{\tau}} \int y(\kindex{s}{ref},t) y(s,t+\tilde{\tau}) \dd t\quad.
\end{equation}
We integrate over the recorded duration covering at least \num{80} cycles of the slowest flapping flags and up to \num{450} cycles for the fastest flapping flags.
The reference position is chosen for all the flags at $\kindex{s}{ref}/L = 0.5$.
The results in \cref{fig:SpatialDeformations}c are shifted such that the first measurable position along $s$ has a zero delay $\tau$ to facilitate the comparison.
The correlation delay increases non-linearly in the first half of the centreline and linearly for the aft part.
The wave speed is extracted from the slope of the linear increase of the correlation delay as a function of $s/L$.
We use the second moment of area as a geometry dependent threshold value to indicate the start of the linear regime as indicated by the shaded area in \cref{fig:SpatialDeformations}c.
The fitting error in this region is of the order of \qty{2}{\percent}, and a sensitivity analysis confirmed $s>\kindex{r}{2}$ as the appropriate region for the fit.  
The total uncertainty on the wave speed is obtained by summing the typical fitting error to the standard deviation of $c$ computed on independent sub-intervals of the recording.

The wave length can now be calculated as $\lambda = c/f$. 
The resulting wave length values for all flags are summarised in \cref{fig:SpatialDeformations}d.
The wavelengths obtained from the spatial correlation approach have been confirmed by direct measurements of the distance between necks and by fitting a propagating wave of amplitude $A(s)$ with $\lambda$, and also a phase lag $\phi$, as fitting parameters to the measured deformations: 
\begin{equation}
y(s,t) = A(s) \sin \left( 2 \pi f t - 2 \pi s / \lambda + \phi \right)\quad.
\end{equation}
The spatial correlation method is faster and more robust than measuring the necks or fitting a propagating wave, and it applies to arbitrary, not sinusoidal waveforms. 

An important conclusion we draw from \cref{fig:SpatialDeformations}d is that the wavelength does not vary with aspect ratio when normalised by the length of the flag.
By averaging over all tested flags, we find an approximately constant value of the dimensionless wavelength of $\lambda/L = \num[separate-uncertainty]{1.87(0.08)}$, which is larger than the wavelength of the second cantilevered beam mode shape ($m=\num{2}$ in \cref{fig:SpatialDeformations}d), but significantly smaller than the first mode ($m=\num{1}$).
The lowest mass ratio series have slightly lower wavelengths.
Low mass ratio flags were characterised by a larger contribution of third mode shape (\cref{fig:EnvelopeAndBeamModes}i).
The wavelength of the third mode is smaller than the wavelength of the first and second mode (\cref{fig:SpatialDeformations}d) and its larger contribution in low mass ratio flags explains the decrease of wavelength with decreasing mass ratio.

The main spatial characteristics of the post-critical flag flapping are thus the amplitude envelope, the tip amplitude, and the wavelength of the travelling deformation wave. 
All flags tested in this study show the same double-neck flapping mode in the post-critical regime.
The spatial characteristics non-dimensionalised by the length of the flag are independent of the aspect ratio, and are predominantly governed by the mass ratio. 
An increase in the mass ratio leads to a shift in the necks toward the tip and a decrease in the nondimensional tip amplitude. 
The dimensionless wavelength remains approximately constant and only slightly decreases for the lowest mass ratios tested.

Next, we will focus on how the aspect ratio and mass ratio affect the temporal characteristics of the deformation waves, including the wave speed, oscillation frequency, and Strouhal numbers.

\subsection{Temporal characteristics of the deformation waves}
\label{subsec:timescales}

\begin{figure}
\includegraphics[width=\linewidth]{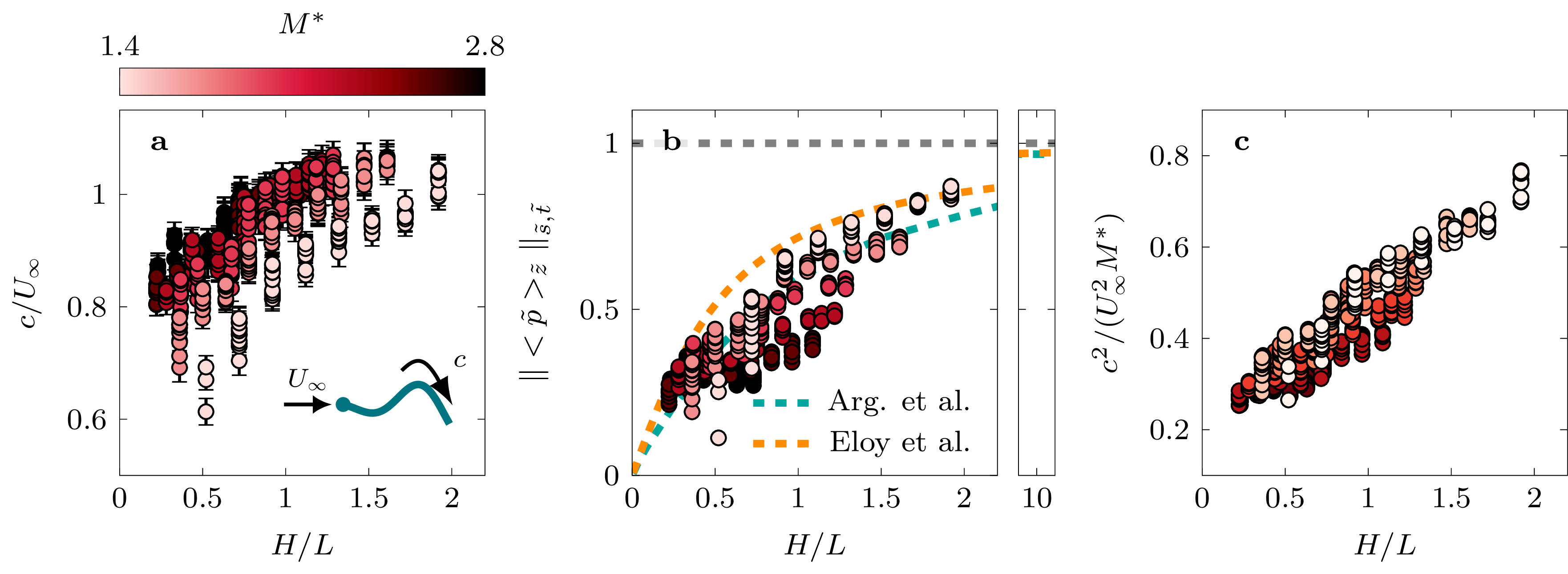}
\caption{(a) Dimensionless wave speed $c/\Uinf$ obtained from the correlation delay plots against aspect ratio $H/L$, coloured by mass ratio $M^*$. 
(b) Estimate of the overall dynamic pressure difference $\|<\tilde{p}>_{\tilde{z}}\|_{\tilde{s},\tilde{t}}$ against aspect ratio compared with the correction factor $A(H/L)$ from \cite{argentina_fluid-flow-induced_2005} and the spanwise averaged pressure correction ${<\bar{p}_n>}_{\bar{z}}$ from \cite{eloy_flutter_2007}. 
(c) Square of the mass ratio corrected dimensionless wave speed as a function of the aspect ratio, coloured by mass ratio.
}
\label{fig:ARTimingsA}	
\end{figure}

The results of the extracted wave speeds from the spatial-correlation delay plots (\cref{fig:SpatialDeformations}c) are presented in \cref{fig:ARTimingsA}a for flags of different aspect ratio, at different mass ratios, and reduced velocities in the post-critical regime ($18<U^*<22$).
The dimensionless wave speed is of the order of \num{1}, with a minimum value of $c/\Uinf = \num{0.61}$ and a maximum value of $c/\Uinf = \num{1.07}$ for our dataset, indicating that the external flow drives the propagation of the wave along the flag.
For individual flags, the standard deviation of $c/\Uinf$ for measurements at different flow velocities is lower than \qty{5}{\percent}.

For tall flags with $H/L\geq1$, the wave speed approximately equals the free stream velocity. 
Dimensionless wave speeds $c/\Uinf > 1$ are observed because the wave speed is measured along the curvilinear abscissa of the flag and the motion travels faster in curvilinear coordinates than in the streamwise direction.
The dimensionless wave speed decreases both with decreasing aspect ratio below $H/L=1$ and with decreasing mass ratio.
Waves travel slower for flags with smaller height when we fix the mass ratio.
The dimensionless wave speed $c/\Uinf$ starts from \num{1.04} at $H/L=1.92$ and decreases to $c/\Uinf=\num{0.61}$ when the aspect ratio drops to $H/L = 0.5$ for the example case of $M^*=\num{1.4}$.
The wave speed also decreases with decreasing mass ratio when we fix the aspect ratio.
Higher mass ratio flapping flags have less inertia, and resist less to the driving force of the flow.
Analytical models often assume that deformations and the associated flow structures are convected at the same speed as the external flow \citep{argentina_fluid-flow-induced_2005}.
\cite{shelley_heavy_2005} also observed a linear increase of the wave speed with flow velocity. 
They found $c/\Uinf \approx \num{0.66}$, which is on the lower side of our data range.
The value $c/\Uinf \approx \num{0.66}$ was obtained for a flag of aspect ratio $H/L=\num{1.46}$ and mass ratio $M^* = \num{19.8}$, which lies outside our data range ($1.4<M^*<2.8$) and suggests a different flutter regime with a smaller dimensionless wavelength ($\lambda/L \sim 0.5$).

For our dataset, the hypothesis $c=\Uinf$ is valid for light or tall flags, but not for heavy or short flags.
A deflection of the flag starting near the root will be pushed towards the tip by the local dynamic pressure exerted by the air flow on the flag.
As the air will try to follow the path of the least resistance, the dynamic pressure will be reduced for flags with shorter heights, as more of the flow can pass around the top or bottom edges when the flag is deflected.
A lower dynamic pressure results in a lower wave speed.

Here, we explore further this hypothesis that decreasing the aspect ratio leads to a reduction in the dynamic pressure experienced by the flag and that the decrease in dynamic pressure is responsible for slowing down the wave propagation.
From the measured kinematics of the flag, we can directly calculate an estimate of the spanwise averaged pressure difference across the flag, \kindex{<\tilde{p}>}{z}, according to the model proposed by \cite{eloy_flutter_2007}:
\begin{equation}\label{eq:pressurediffcalc}
\kindex{<\tilde{p}>}{z}(\tilde{s},\tilde{t}) = \frac{1}{M^*}\left[
\frac{\partial^2 \tilde{y}(\tilde{s},\tilde{t})}{\partial\tilde{t}^2} 
+ \frac{1}{U^{*2}} \frac{\partial^4 \tilde{y}(\tilde{s},\tilde{t})}{\partial \tilde{x}^4} \right]\quad, 
\end{equation}
where $\kindex{<.>}{z}$ stands for spanwise averaged quantities, and $\tilde{.}$ stands for dimensionless quantities:
\begin{equation}
\tilde{y}=\frac{y}{L}, ~ \tilde{s}=\frac{s}{L}, ~ \tilde{x} = \frac{x}{L}, ~ \tilde{t} = \frac{t\Uinf}{L}, ~ \tilde{p} = \frac{p}{\kindex{\rho}{f} \Uinf^2}\quad.
\end{equation}
\Cref{eq:pressurediffcalc} is the result of the linearised conservation of transverse momentum equation (more details on the derivation are included in appendix \cref{app:wavespeedandpressure}).
To obtain a single representative value of the pressure difference per flapping flag at a given $U^*$, we compute the overall euclidean norm of the calculated $\kindex{<\tilde{p}>}{z} (\tilde{s},\tilde{t})$ over both $\tilde{s}$ and $\tilde{t}$.
We denote this overall dynamic pressure difference by $\|{<\tilde{p}>}_{\tilde{z}}\|_{\tilde{s},\tilde{t}}$ and show the results for our entire dataset in \cref{fig:ARTimingsA}b.
The estimated overall dynamic pressure difference increases with increasing aspect ratio and decreasing mass ratio.
It is more sensitive to variations in aspect ratio than in mass ratio.

Our data is compared in \cref{fig:ARTimingsA}b to predictions by \cite{eloy_flutter_2007} and \cite{argentina_fluid-flow-induced_2005}.
The predictions by \cite{eloy_flutter_2007} stem from a linear stability analysis of the three-dimensional potential flow around a flag of finite height in the Fourier space. 
This approach yields a single spanwise-averaged pressure difference as a function of the normalised streamwise wave number $\kindex{\bar{k}}{x} = \pi H/\lambda$, which they fitted by 
\begin{equation}
{<\bar{p}>}_{\bar{z}} \left(\kindex{\bar{k}}{x}\right)~ \approx ~1 - \frac{1}{\displaystyle 2\pi \frac{L}{\lambda} \frac{H}{L} + \exp{\left(\left(\pi/4-2 \right)\pi \frac{L}{\lambda}\frac{H}{L} \right)}}\quad.
\end{equation}
The fit is shown in \cref{fig:ARTimingsA}b by the dashed line. 
\cite{argentina_fluid-flow-induced_2005} followed a different approach to determine three-dimensional or aspect ratio effect on pressure. 
They numerically calculated a correction factor for the pressure $A$, which is the ratio of the spanwise-averaged non-circulatory potential of a plate of finite height and the non-circulatory potential of an infinite-height plate facing a constant flow.
We extract the values of $A(H/L)$ from figure 5b in \cite{argentina_fluid-flow-induced_2005} and include it in \cref{fig:ARTimingsA}b for further comparison.
Our experimental data is in good agreement with the theoretical prediction by \cite{eloy_flutter_2007} and the numerically obtained correction factor by \cite{argentina_fluid-flow-induced_2005}.
Our pressure estimates lie between the values predicted by both models for low and high aspect ratios and are smaller in the intermediate region $0.5<H/L<1.5$.
The agreement between the two pressure models and our data support the argument that edge effects induce a reduction in the overall pressure difference during flapping.

Now, we want to link the pressure to the wave speed.
Based on the comparison of the order of magnitude of the terms in the momentum equation, which we discuss in \cref{app:wavespeedandpressure}, we obtain the following scaling for the dimensionless wave speed:  
\begin{equation}\label{eq:mainwavespeedscaling}
c/\Uinf \propto \sqrt{M^* {<\tilde{p}>}_{\tilde{z}}}\quad.
\end{equation}
This scaling indicates that there is a contribution of the mass ratio and the dynamic pressure difference on the wave speed.
This scaling confirms our experimental observation of increasing wave speed with increasing mass ratio (\cref{fig:ARTimingsA}a) and with increasing dynamic pressure difference, which occurs when the aspect ratio increases (\cref{fig:ARTimingsA}b).
  
Based on \cref{eq:mainwavespeedscaling}, we can introduce the mass ratio corrected dimensionless wave speed $c/(\Uinf \sqrt{M^*})$, which is directly proportional to the square root of the norm of the pressure difference within a flutter mode (\cref{fig:wavespeedvspressure}a-b in the appendix).
The mass ratio corrected dimensionless wave speed varies only with the aspect ratio (\cref{fig:ARTimingsA}c) and links the aerodynamic forcing that depends on the geometric property of the flag to the spatial and temporal parameters characterising the flapping response.

\begin{figure}
\includegraphics[width=\linewidth]{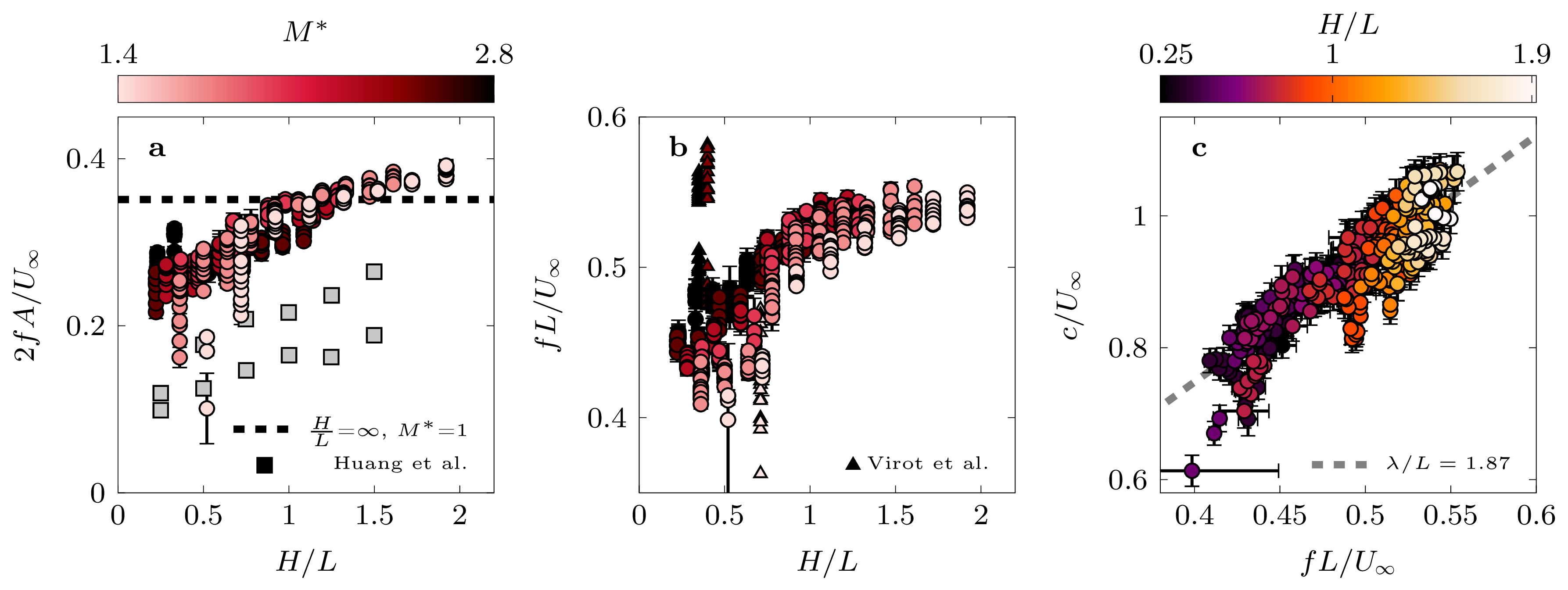}
\caption{(a) Experimental values of the amplitude-based Strouhal number $\kindex{St}{A} = 2fA/\Uinf$ (circles) as a function of aspect ratio. 
For comparison, we included results from three-dimensional numerical simulations at $Re=500$ (rectangles) and the infinite aspect-ratio limit (dashed line) from \cite{huang_three-dimensional_2010}. 
(b) Experimental values of the length-based Strouhal number $fL/\Uinf$ (circles) as a function of aspect ratio. 
For comparison, we included experimental results from \cite{virot_fluttering_2013} (triangles). 
(c) Dimensionless wave speed $c/\Uinf$ against flapping frequency $fL/\Uinf$.
The dashed line indicates the linear relationship between $c/\Uinf$ and $fL/\Uinf$ for the mean measured dimensionless wavelength $\lambda/L=1.87$.}
\label{fig:ARTimingsB}	
\end{figure}

The dimensional wave propagation speed is directly linked to the flapping frequency through $c=\lambda f$.
The flapping frequency is typically non-dimensionalised using a characteristic length scale and the flow velocity to form a Strouhal number.
In literature, both the tip to tip amplitude $2A$ and the flag length $L$ are used to define a Strouhal number.
In \cref{fig:ARTimingsB}a, we show the amplitude-based Strouhal number $\kindex{St}{A} = 2fA/\Uinf$ against aspect ratio for all our data. 
The amplitude-based Strouhal number increases from \numrange{0.10}{0.39} with increasing aspect ratio, with most of the data points above \num{0.20}.
We compare our data in \cref{fig:ARTimingsB}a to results from three-dimensional numerical simulations by \cite{huang_three-dimensional_2010}, who also observed an increase of $\kindex{St}{A}$ with aspect ratio.
The values of $\kindex{St}{A}$ from the simulations are lower in magnitude due to a different flapping regime ($M^* = \num{1}$ and \num{0.25} corresponding to the single neck flutter).

In \cref{fig:ARTimingsB}b, we show the length-based Strouhal number $\kindex{St}{L} = fL/\Uinf$ against aspect ratio for all our data.
The dimensionless tip frequency $fL/\Uinf$ increases with aspect ratio, and to a lesser extent with mass ratio for our tested flags, which are all in the double-neck flutter regime.
On the lower side, $fL/\Uinf = \num{0.40}$ and a half stroke occurs in \num{1.25} convective times ($L/\Uinf$).
On the higher side, $fL/\Uinf = \num{0.55}$ such that half a stroke of the tip occurs in only \num{0.91} convective times.
Experimental values of $fL/\Uinf \sim 0.4-0.6$ are reported in \cite{virot_fluttering_2013} for a two-neck flutter, and most of the references in the literature report $fL/\Uinf \approx 0.23$ for single neck flutter.
At constant reduced velocity and within the single-neck flutter regime, \cite{chun2010flutter} report a slight increase of the dimensionless flapping frequency with aspect ratio, from $\kindex{St}{L}=\num{0.22}$ to \num{0.25} at $U^* = \num{15}$ when increasing the aspect ratio from \num{0.5} to \num{1.24}.
\cite{eloy_aeroelastic_2008} report $\kindex{St}{L} = \num{0.59}$ and \num{0.40} for two example flags flapping in the double-neck ($M^*=1.94$) and single neck ($M^*=0.74$) regime respectively, and they show that linear models can predict the flapping frequency at onset with an accuracy of \qtyrange{23}{32}{\percent}.
Overall, the magnitude of the dimensionless tip-frequency depends on the flutter mode, and increases with aspect ratio within a flutter regime.

The relationship between tip frequency and wave speed can be rewritten as 
\begin{equation} 
\frac{c}{\Uinf} = \frac{\lambda}{L} \times \frac{fL}{\Uinf}.
\end{equation}
In \cref{fig:ARTimingsB}c, we plot the measured wave speed against the measured tip frequency.
For our dataset, $c/\Uinf$ indeed increases approximately linearly with $fL/\Uinf$, with a slope equal to $\lambda/L$.
The range of length-based Strouhal number within a given flutter regime is explained by variations in the wave propagation speed.
Larger variations in the length-based Strouhal number are associated with variations in the flutter regime.
For a flutter regime with more necks and a lower wavelength, we expect higher length-based Strouhal numbers.

\subsection{Effect of aspect ratio on wake quantities}
\label{subsec:wakequantities}

\begin{figure}
\includegraphics[width=\linewidth]{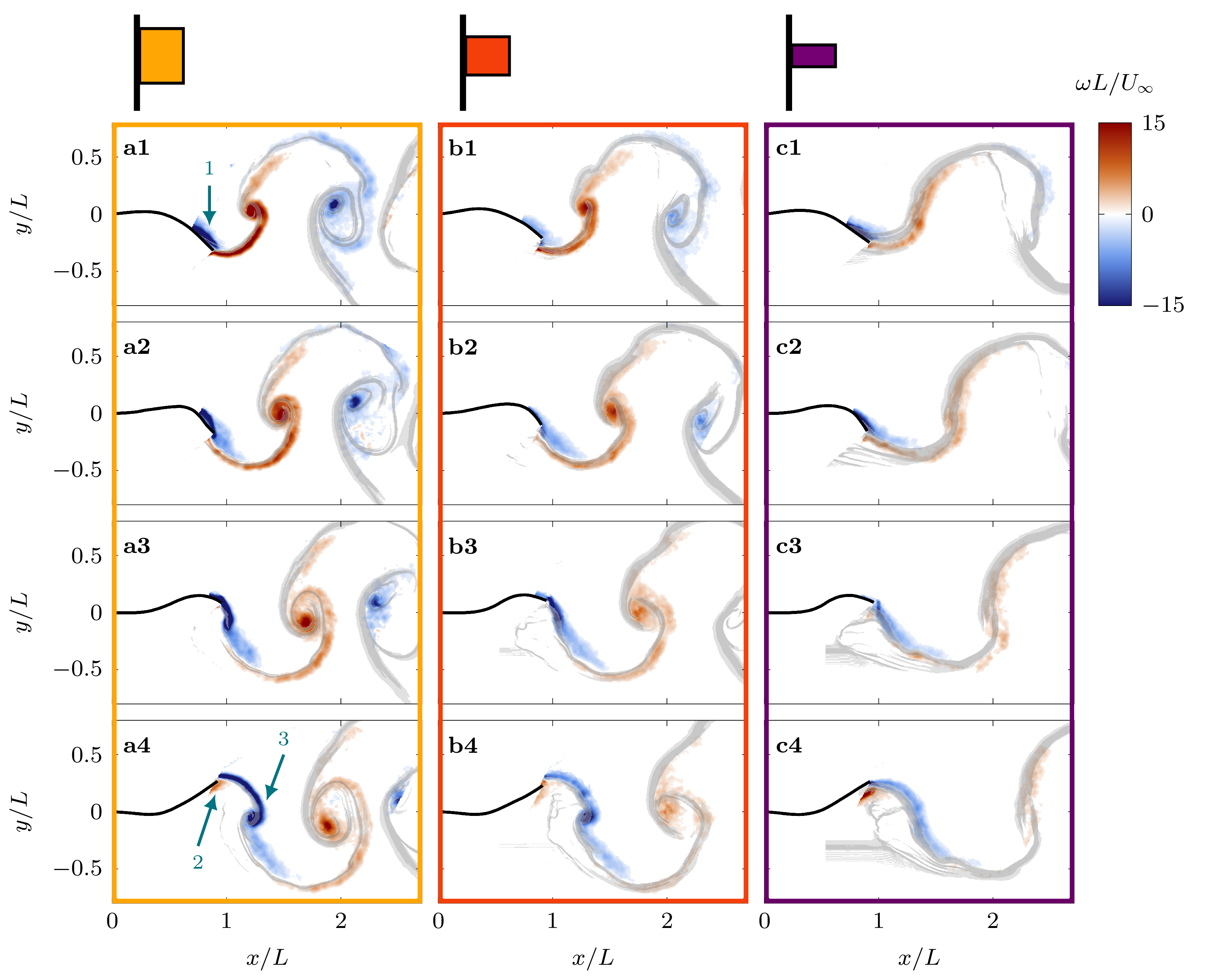}
\caption{Snapshots of phase-averaged dimensionless vorticity $\omega L/\Uinf$ for three flags with different aspect ratio: (a) $H/L = \num{1.25}$, (b) $H/L = \num{0.875}$, (c) $H/L = \num{0.5}$. 
The three flags have the same length and are tested at the same flow velocity $\Uinf=\qty{7.5}{\meter\per\second}$.
The selected snapshots correspond to four instants during the upstroke: (1) $t/T=0$ when the tip deflection $\kindex{y}{tip}$ reaches its minimum value, (2) $t/T=0.16$, (3) $t/T=0.32$ shortly after the tip passes the midline and (4) $t/T=0.48$. 
Ridges of the negative finite time Lyapunov exponent are overlaid on the vorticity field.}
\label{fig:VorticitySnapshots}
\end{figure}

To characterise how the vortex formation varies with aspect ratio, we performed flow measurements at mid-height for a series of flags with the same mass ratio $M^*=\num{2.27}$.
In \cref{fig:VorticitySnapshots}, we show snapshots of vorticity for three different flags with different aspect ratios ($H/L=\num{1.25}$, \num{0.875} and \num{0.5} for \cref{fig:VorticitySnapshots}).
For each flag, we selected four snapshots covering half a flapping cycle.
We focus on the upward stroke of the flag, where the tip moves from negative $y$ to positive $y$, which respectively correspond to the bottom and the top in \cref{fig:VorticitySnapshots}.
The upward stroke begins when the tip is at the most extreme deflection (first row of \cref{fig:VorticitySnapshots}).
Negative vorticity is present along the outside of the curvature of the flag close to the tip (indicated by arrow 1 in \cref{fig:VorticitySnapshots}a1).
When the flag starts to move back toward the centerline, the negative vorticity is continuously released into the wake (second and third row of \cref{fig:VorticitySnapshots}).
When the flag reaches its maximum deflection on the other side, there is no more negative vorticity released at the tip (fourth row of \cref{fig:VorticitySnapshots}).
Before switching to the downward motion, positive vorticity accumulates near the tip (indicated by arrow 2 \cref{fig:VorticitySnapshots}a4) and that phase corresponds to the start of the second half cycle.
For the three different aspect ratio flags, negative and positive vorticity are alternatingly released in the wake during upstroke and downstroke.
For the tallest flag, the vorticity released in the wake rolls up into a coherent concentrated vortex core.
With decreasing aspect ratio, the intensity of the vorticity in the wake decreases and the wake vortices are less coherent.

We highlight the topology of the vortex street in the wake using the negative finite time Lyapunov exponent (nFTLE) \citep{green_unsteady_2011}.
The nFTLE ridges are overlaid on the vorticity fields to depict the attractor lines in the flow (\cref{fig:VorticitySnapshots}). 
The nFTLE ridges coincide with regions of increased shear.
As the vortex street develops in the wake, the nFTLE ridges undulate in streamwise direction.
The wavelength of the nFTLE ridges in the streamwise direction decreases with increasing aspect ratio.
The wavelength of the undulating nFTLE ridge is approximately twice the distance between two consecutive vortex cores, which is approximately $\Uinf/2f$ if we assume that the vortices are convected downstream with a velocity equal to $\Uinf$.
As taller flags flap faster, the wake wavelength $\sim \Uinf/f$ decreases and consecutive vortices are shed closer to each other.
The nFTLE ridge rolls-up in the close wake ($1<x/L<2$) and a coherent vortex core is convected downstream for the tall flag (indicated by arrow 3 in \cref{fig:VorticitySnapshots}a).
No vortex roll-up is observed for the shortest flag and the vorticity is more distributed along the shear layer (\cref{fig:VorticitySnapshots}c).

\begin{figure}
\includegraphics[width=\linewidth]{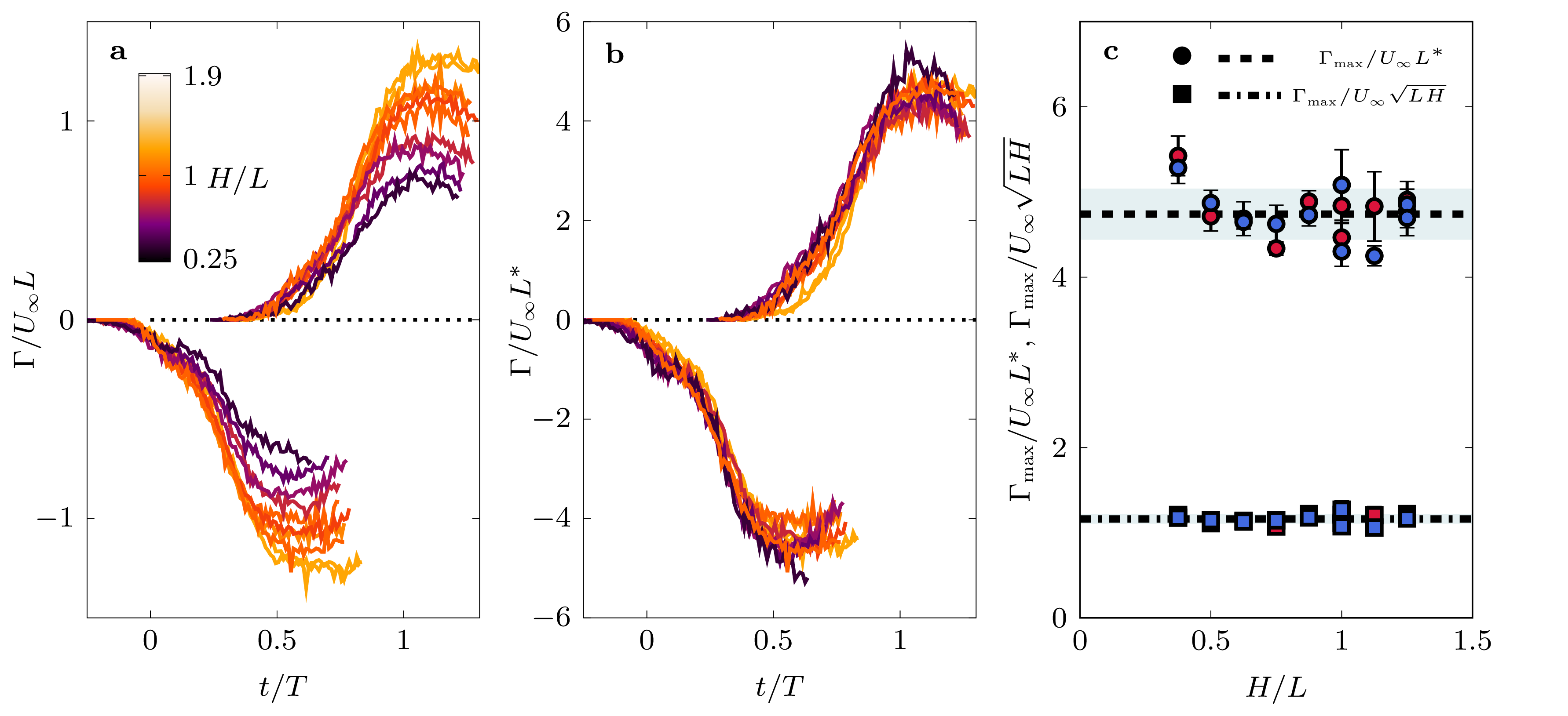}
\caption{Evolution of the phase-averaged (a) dimensionless circulation $\Gamma/\Uinf L$  and (b) the scaled circulation $\Gamma/\Uinf L^*$ as a function of dimensionless time $t/T$. 
The instants $t/T=\num{0}$ and $\num{0.5}$ correspond to the moment of minimum and maximum tip deflection. 
(c) Maxima of the positive and negative circulation using $L^*$ and $\sqrt{LH}$ for the aspect ratio scaling.
The height of the shaded areas indicates the standard deviation for each scaling of $\kindex{\Gamma}{max}$}
\label{fig:circulation}	
\end{figure}

To quantify how the intensity of vorticity released in the wake increases with increasing aspect ratio, we calculate the circulation released from the tip during half a stroke. 
The three-dimensional topology of the vortices in the wake is described as a horseshoe-shaped vortex \citep{huang_three-dimensional_2010}. 
Since circulation is conserved along a vortex tube, mid-plane flow data are relevant to describe the near wake dynamics behind a flapping flag. 
The technical details for the integration of the circulation is described in \cref{app:circulation} of the appendix.
The dimensionless positive and negative circulation $\Gamma/\Uinf L$ is presented as a function of the dimensionless time $t/T$ with $t=0$ corresponding to the maximum downward deflection of the tip.
The sharp increase of negative circulation occurs shortly after the upward motion of the tip starts and negative vorticity is released in the wake.
Vorticity release changes sign when the tip reverts at the end of the upstroke, no more negative vorticity is released.
The existing negative vorticity is convected downstream, and negative circulation reaches a plateau after the tip has started the downward motion ($t/T > 0.5$ in \cref{fig:circulation}a).
At the tip reversal around $t/T=0.5$, positive vorticity is released, and we observe the same evolution of the circulation as in the first half cycle, but with opposite sign. 
For both positive and negative circulation, the maximum value occurs around or slightly after the tip reversal.
Maxima of circulation nearly double from $\kindex{\Gamma}{max}/\Uinf L= \num{0.72}$ to \num{1.35} with aspect ratios increasing from $H/L = \num{0.375}$ to \num{1.25} (\cref{fig:circulation}a).
The variations of the non-dimensional circulation we observe for flags of the same length, reduced velocity, and Reynolds number suggest that $L$ is not the only characteristic dimension to scale the flow quantities.
We propose two alternative length scales to account for the variations in circulation: the ratio of the area over the perimeter $L^* = LH/(2L+2H)$ and the square root of the area $\sqrt{LH}$.
The area-to-perimeter ratio $L^*$ is a fourth of the hydraulic diameter which is used to scale the wake dynamics behind flat rectangular plates \citep{fernando2016vortex} and flat elliptical plates \citep{fernando_reynolds-number_2016}.
In the context of vortex formation behind an accelerated plate, the total mass displaced scales with the area of the moving object, and the vorticity is generated along the perimeter, such that the area-to-perimeter ratio is expected to be a measure for the reciprocal of the vorticity production per mass displaced. 
Alternatively, the wake behind a plate is expected to scale with the equivalent diameter, i.e. the size of a circle of equal area as the plate \citep{higuchi1996three}, which is the equivalent to using $\sqrt{LH}$.
The square root of area $\sqrt{LH}$ performed better at scaling circulation, hydrodynamic impulse, and kinetic energy for accelerated polygonal disks \citep{caverly_invariant_2025}.
The square root of area is also more robust in the case of fractal objects in steady motions where the perimeter to area ratio can become extremely large without having a measurable influence on the flow properties \citep{nedic_drag_2013}.
Both length scales, $L^*$ and $\sqrt{LH}$, have been successful at collapsing circulation data for plates of different geometry and aspect ratio and are good candidates to account for three-dimensional effects in the wake of flapping flags.

We observe a collapse of the temporal evolution of the circulation for the different aspect ratios when replacing $L$ by $L^* = LH/(2L+2H)$ (\cref{fig:circulation}b). 
The same holds when we replace $L$ by $\sqrt{LH}$. 
For different aspect ratios, we obtain a similar mean maximum value of  $\kindex{\Gamma^*}{max} = \kindex{\Gamma}{max}/\Uinf L^* = \num[separate-uncertainty]{4.74(0.30)}$ and $\kindex{\Gamma^\dagger}{max} = \kindex{\Gamma}{max}/\Uinf \sqrt{LH} = \num[separate-uncertainty]{1.16(0.06)}$ (\cref{fig:circulation}c).
The standard deviations are \qty{7}{\percent} and \qty{5}{\percent}, for the two scaling approaches respectively.
The collapse of the different curves shows that the right length scale for the mid-plane circulation should include the height to account for three-dimensional effects.

\subsection{Link between mean drag coefficient and flapping dynamics}
\label{subsec:dragscaling}

\begin{figure}
\centering
\includegraphics[width=\linewidth]{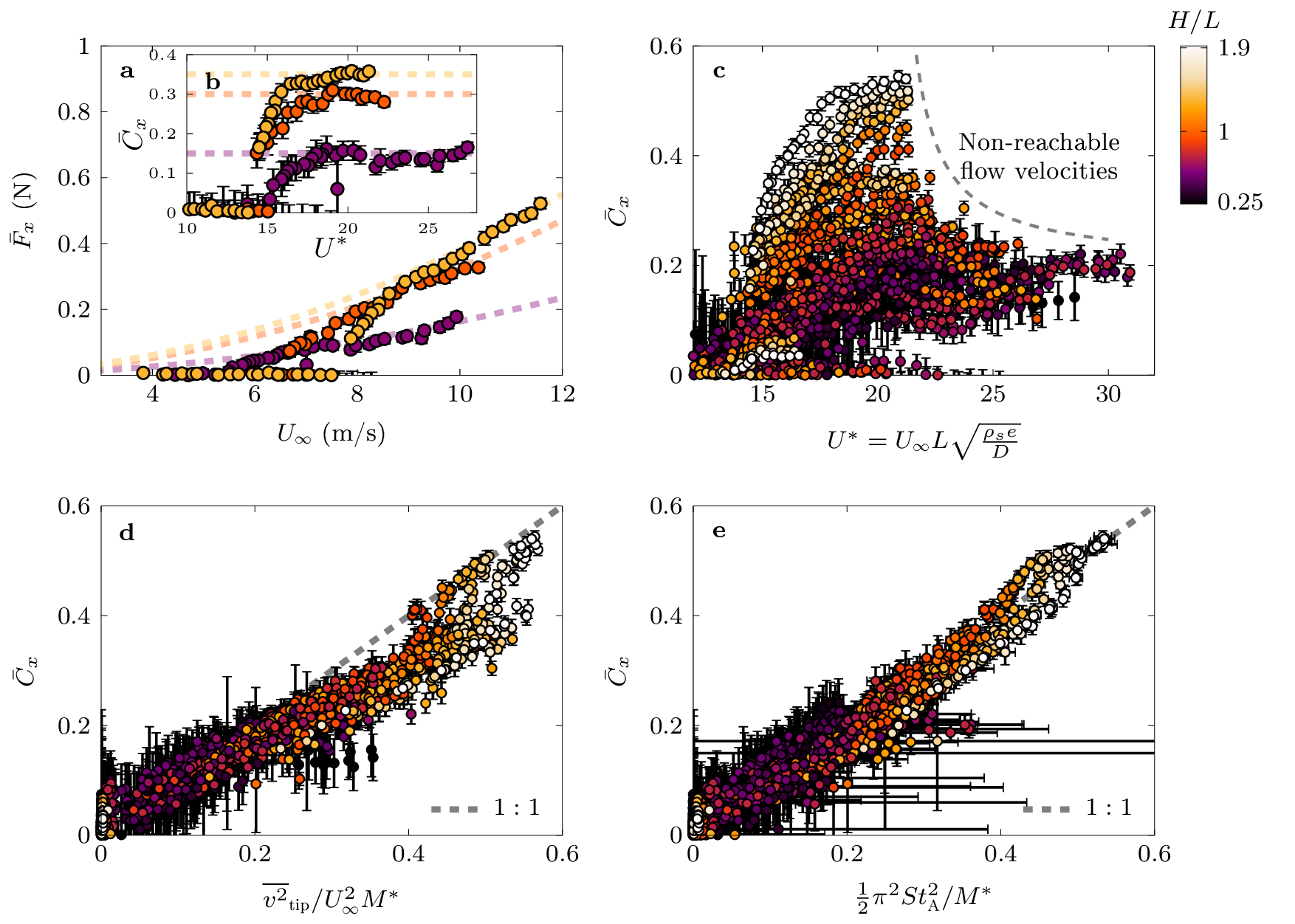}
\caption{(a) Mean drag as function of flow velocity $\Uinf$ for three flags of equal area $HL = \qty{192}{cm\squared}$ and aspect ratio $H/L = \numlist{0.59;0.98;1.33}$. 
(b) Corresponding mean drag coefficient $\kindex{\bar{C}}{x}$. 
(c) Mean drag coefficient $\kindex{\bar{C}}{x}$ for flags with $1.4<M^*<2.8$ and $0.22 < H/L < 1.92$. 
The empty area in the top right part corresponds to experimental conditions that the set-up can not reach. 
Scaling of the mean drag coefficient with (d) $\kindex{\overline{v^2}}{tip} / \Uinf^2 M^*$ and (e) $\frac{1}{2}\pi^2 \kindex{St}{A}^2/M^*$.}
\label{fig:drag}	
\end{figure}

The mean drag force $\kindex{\bar{F}}{x}$, which is the time-average force in the streamwise direction, at the root of the flag is measured systematically for different flow velocities.
Positive values of $\kindex{\bar{F}}{x}$ indicate a net drag force experienced by the flag. 
We take the example of three flags of equal area to analyse how the drag varies with flow velocity (\cref{fig:drag}a).
The mean drag is close to zero before the flapping onset as the only contribution to drag comes from the friction drag.
A jump to a significant positive mean drag value occurs after the flapping onset when the pressure drag increase due to the flapping.
With increasing flow velocity, the mean drag increases following a quadratic increase with flow velocity as we expect for bluff bodies (\cref{fig:drag}a). 
The mean drag significantly differs between the three flags of same area when compared at the same flow velocity.
The three flags have different lengths, so their flapping onset occurs at a different flow velocity. 
The short flag starts flapping before the tall flag such that the drag of the short flag is higher around $\Uinf = \qty{7}{\meter\per\second}$.
When the tallest flag finally starts flapping, the mean drag increases faster with $\Uinf$ than for the other flags and overtakes the values of the other flags for $\Uinf > \qty{8}{\meter\per\second}$.
The mean drag coefficient 
\begin{equation}
\kindex{\bar{C}}{x} = \frac{\kindex{\bar{F}}{x}}{\frac{1}{2}\kindex{\rho}{f} H L \Uinf^2}
\end{equation} 
of the three example flags undergoes a transition from \num{0} to a plateau at post-critical velocities. 
The plateau values vary between \num{0.15} for $H/L = \num{0.59}$ to \num{0.35} for $H/L = \num{1.33}$ (\cref{fig:drag}b).
The individual values of mean drag coefficient for the \num{48} flags tested display a wide distribution ranging from $\kindex{\bar{C}}{x} = \numrange{0}{0.55}$ with no clear relationship with the reduced velocity $U^*$ (\cref{fig:drag}c).
At fixed reduced velocity $U^*$, the mean drag coefficient $\kindex{\bar{C}}{x}$ tends to increase with aspect ratio.
The scattering of mean drag coefficient suggests that the typical aerodynamic force normalisation with the surface area and the square of the flow velocity fails to scale the mean drag of flapping flags.
The projected area $2AH$, which takes into account the difference in wake width, is not a suitable replacement for the flag area $HL$ as shown in the appendix (\cref{appForceData}).
The non-dimensionalisation of the drag coefficient using the projected area does not reduce the overall scattering of the drag coefficient values.

To explain the distribution of mean drag coefficient, we revisit the work of \cite{thoma_schlenkernde_1939} and derive an empirical prediction using the measured flag tip velocity.
\cite{thoma_schlenkernde_1939} presented a model for a fully flexible rope oscillating under the action of normal forces and derived a two-dimensional estimation of the mean tension at the root of the rope.
The mean drag per unit height was derived as
\begin{equation}\label{eq:FxThoma1939}
\kindex{\bar{F}}{x} = \frac{1}{2} m \kindex{\overline{v^2}}{tip}\quad,
\end{equation}
where $\overline{.}$ stands for a time-averaged quantity, and $m = \kindex{\rho}{s} e H$ is the mass per unit length of the rope.
For flags, $\kindex{\rho}{s} e H L$ is the total mass.
The right-hand side of \cref{eq:FxThoma1939} represents the tension induced by the in-plane kinetic energy of the flag.
In dimensionless quantities, \cref{eq:FxThoma1939} can be formulated as
\begin{equation}\label{eq:CxTipVelocityScalingStep1}
\kindex{\bar{C}}{x} = \displaystyle\frac{\frac{1}{2} \kindex{\rho}{s} e H \kindex{\overline{v^2}}{tip}}{\frac{1}{2} \kindex{\rho}{f} HL \Uinf^2 } = \frac{ \kindex{\rho}{s} e }{ \kindex{\rho}{f} L } \times \frac{\kindex{\overline{v^2}}{tip}}{\Uinf^2}\quad,
\end{equation}
which can be rewritten as
\begin{equation}
\label{eq:CxTipVelocityScaling}
\kindex{\bar{C}}{x} = \frac{1}{M^*}  \frac{\kindex{\overline{v^2}}{tip}}{\Uinf^2}\quad.
\end{equation}
The mean drag coefficient is proportional to the ratio of the mean square tip velocity and the incoming flow velocity squared with the mass ratio as the proportionality factor.
In \cref{fig:drag}d, we combine the deformation measurement $\kindex{v}{tip}$ and the load cell data $\kindex{\bar{C}}{x}$.
The two-dimensional kinematic model of \cref{eq:CxTipVelocityScaling} describes our data well. 
The influence of the aspect ratio on the mean drag coefficient is indirectly included in the reduced tip speed.

Following \cite{moretti_tension_2003}, we can simplify the tip speed by assuming that the tip displacement is described by $\kindex{y}{tip}=-A \cos{2\pi f t}$.
The simplified tip velocity takes the form $\overline{\kindex{v}{tip}^2}=2 \pi^2 f^2 A^2$.
The mean drag coefficient estimate can be rewritten as
\begin{equation}
\label{eq:StABasedDragEstimate}
\kindex{\bar{C}}{x} = \frac{\pi^2}{2} \times \frac{\kindex{St}{A}^2}{M^*}\quad,
\end{equation}
where the amplitude-based Strouhal number $\kindex{St}{A}= 2fA/\Uinf$, shown in \cref{fig:ARTimingsB}a, accounts for the increasing tip speed with increasing aspect ratio.
In \cref{fig:drag}e, we show the measured mean drag coefficient against the simplified estimate based on $\kindex{St}{A}$ from \cref{eq:StABasedDragEstimate}.
The average discrepancies between the model and the measured data $\left| \kindex{\bar{C}}{x}  - (\pi^2/2)\times\kindex{St}{A}^2/M^*\right|$ is \num{0.023}.
The simplified empirical model is more accurate than the empirical model that directly uses the tip velocity (\cref{eq:CxTipVelocityScaling}) which has an average discrepancy $\left| \kindex{\bar{C}}{x}  - \kindex{v}{tip}^2/\Uinf^2 M^*\right|$ of \num{0.039}.
Overall, an increase in aspect ratio induces an increase in tip speed (\cref{fig:drag}d), an increase in amplitude-based Strouhal number (\cref{fig:drag}e), and an increase in mean drag coefficient.

The models based on the tip velocity or the amplitude-based Strouhal number are empirical in nature.  
They indicate the strong physical connection between the measured tip-based quantities and the drag coefficient, but they do not provide an a priori prediction based on reduced velocity, mass ratio, and aspect ratio. 
The reduced velocity governs whether we are in the no-flapping regime, where the drag is dominated by the friction drag, the fully established flapping regime, where the drag increases quadratically with the free-stream velocity, or in the transition regime between both (\cref{fig:drag}a).
In the fully established flapping regime, the mass ratio mainly governs the flapping mode, which is associated with a constant wave length.
As the wave length connects the wave speed and the flapping frequency (\cref{fig:ARTimingsB}c), the final requirement for a fully predictive model, would be the functional relationship of the wave speed as a function of the aspect ratio. 
Potential avenues to obtain such a relationship could be based on the unsteady Prandtl lifting line theory which could aid to describe how the aspect ratio affects the spanwise loading, and subsequently the vortex strength, and mean drag force \citep{bird_unsteady_2021, bird_usefulness_2022,guermond_unified_1991, dimitriadis_unsteady_2024}.

\section{Conclusion}

We report systematic experimental measurements of rectangular flags in the post-critical flapping regime.
Finite height flags, with lower aspect ratio, display different kinematics than quasi two-dimensional flags of large aspect ratio. 
The dimensionless flapping frequency, wave speed, and mean drag coefficient decrease with decreasing aspect ratio.
We present a causal chain that connects the geometric flag properties ($H/L$ and $M^*$) to the spatial and temporal characteristics of the flapping dynamics and explains the spread of flapping frequency and drag values reported in the literature.

The flow around the top and bottom edges of finite height flags reduces the dynamic pressure difference across the flag. 
We introduce the mass ratio corrected dimensionless wave speed, which is directly linked to the square root of the pressure difference, and show that it varies only with the aspect ratio.
It links the aerodynamic forcing that depends on the geometric property of the flag to the spatial and temporal parameters characterising the flapping response.

Among those parameters are the wave length and the wave speed of the deformation wave. 
There parameters have been mostly overlooked in previous studies, and we highlight here the crucial role they play in the mechanistic explanation of the flapping dynamics.
We show that the dimensionless wave length $\lambda/L$ is constant within a given flapping regime, and we use the relationship $c=\lambda f$ to explain the lower flapping frequency and slower tip speed observed in lower aspect ratio flags.
With lower tip speed and stronger edge effects, less vorticity is released in the wake of slender flags and vortices do not roll up any more.
The typical length scale to predict the reduction in released circulation during half a stroke should include the height to account for the three-dimensional nature of the wake.
The ratio of the area over the perimeter $L^*$, and the square root of area $\sqrt{LH}$ include both the information of the height and the length and are suitable scale parameters for the circulation of different aspect ratio flags.
The values of mean drag coefficient measured for flags of different aspect ratios and mass ratios range widely from \numrange{0}{0.55}, confirming the enigma that already puzzled \cite{taneda_waving_1968}.
We show that the evolution of drag with flow velocity follows three regimes depending on its flapping state: almost no drag in the stable regime, a transition phase with a linear increase of drag with $\Uinf$, and a quadratic regime with constant mean drag coefficient.
The planar flag surface area, or the projected area are not sufficient to explain the spread of measured mean drag coefficient.
By revisiting the model proposed by \cite{thoma_schlenkernde_1939}, we show that the characteristic velocity scale for the drag is the tip velocity.
This empirical model also shows that the reduction of tip frequency with decreasing aspect ratio, which we explained using the mass ratio corrected dimensionless wave speed, drives the reduction of mean drag coefficient for shorter flags.

This paper provides a mechanistic explanation connecting the deformation, flow, and force characteristics of flapping flags of different aspect ratio in the post-critical regime.
Systematic high temporally and spatially resolved deformation data, unlocked by event-based imaging, give access to precise extraction of wave propagation speeds, wave lengths, and tip speeds for a variety of flag aspect ratios, mass ratios, and reduced velocities.
Simultaneous deformation and force measurements for \num{48} different flags at various reduced velocities allowed us to validate analytical drag estimations based on the tip velocity. 
Future work could investigate the relationship between time-varying forces, three-dimensional flow fields, and deformations of more complex flag shapes.


\backsection[Acknowledgements]{
GR wishes to acknowledge the support from Alexandros Anastasiadis, Sahar Rezapour and Lorenzo Ermanni.
}

\backsection[Funding]{
This work was supported by the Swiss National Science Foundation under grant number 200021-232147.
}

\backsection[Declaration of interests]{The authors report no conflict of interest.}




\appendix

\section{Beam mode}\label{app:beammodes}
\begin{figure}
\includegraphics{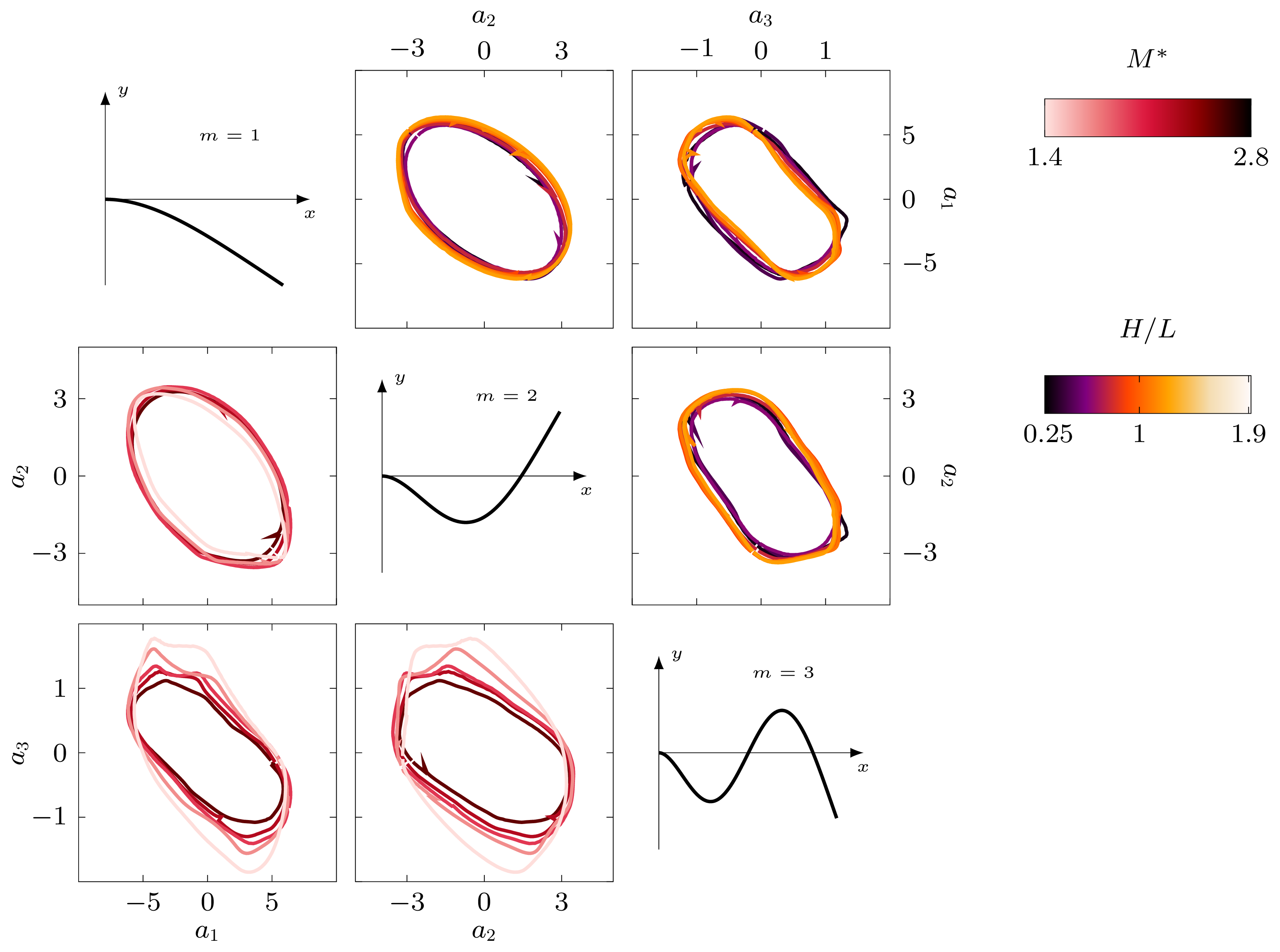}
\caption{Details on the phase-averaged mode coefficients corresponding to the three first cantilevered beam mode shapes. In the top right panels, we summarise results of flags of the same mass ratio $M^*=\num{2.27}$, the same flags as in \cref{fig:EnvelopeAndBeamModes}h. 
In the bottom left panels, we summarise results of flags of similar aspect ratio $H/L \approx 1$, the same flags as in \cref{fig:EnvelopeAndBeamModes}i.}
\label{fig:appCantileveredModes}
\end{figure}

The phase-averaged coefficients of the cantilevered modes are computed from the projection of the dimensionless transverse deformation $y(s,t)/L$ onto the beam mode shapes $\kindex{\psi}{i}(s)$:
\begin{equation}
	\kindex{a}{i}(t) = \frac{1}{L} \int_0^L  \frac{y(s,t)}{L} \kindex{\psi}{i}(s) ds.
\end{equation}
The three first cantilevered mode shapes $\kindex{\psi}{i}$ are plotted on the diagonal and the limit cycles for flags of different mass ratio or different aspect ratio are plotted on the different panels of the matrix in \cref{fig:appCantileveredModes}.
We use the cantilever beam modes here as they form a common and well-known modal base for decomposing cantilevered deformations. The modes are described by analytical expressions, which facilitates comparison between different studies.
An alternative decomposition using, for example, a proper orthogonal decomposition of the measured flag deformation, would yield the same conclusions, but the modes would be data-driven and could vary between data sets.

\section{Pressure estimates}\label{app:wavespeedandpressure}

In this section, we describe the computation of the pressure estimate and we derive a relationship between pressure and wave speed.
We follow the model from \cite{eloy_flutter_2007}.
This model is linear and two-dimensional in terms of deformations, it assumes that the motion of the centreline $y$ is small ($y \ll L $) and that it does not vary with $z$.
The fully three-dimensional plate model according to \cite{huang_three-dimensional_2010} considers invariance along the span ($\partial / \partial z = 0$).
As a consequence, the spatial derivatives of the displacement along the span $z$ that appear in the bending and twisting terms would cancel, and we would recover the same two-dimensional momentum equations as given by the beam equations.
The model also assumes that the motion of the flag is only driven by normal pressure forces.
With spanwise invariance of the motion, the local momentum equation is averaged over the span $-H/2 < z < H/2$.
The spanwise-averaged momentum balance of the linear model writes as
\begin{equation}
	\label{eq:momentumbalance}
	\frac{\partial^2 \tilde{y}}{\partial\tilde{t}^2} + \frac{1}{U^{*2}} \frac{\partial^4 \tilde{y}}{\partial \tilde{x}^4} = M^* {<\tilde{p}>}_{\tilde{z}},
\end{equation}
where the $\tilde{~}$ stands for dimensionless quantities normalised by $L$, $\Uinf$ and $\kindex{\rho}{f}$:
\begin{equation}
	\tilde{y}=\frac{y}{L}, ~ \tilde{x} = \frac{x}{L}, ~ \tilde{t} = \frac{t\Uinf}{L}, ~ \tilde{p} = \frac{p}{\kindex{\rho}{f} \Uinf^2}.
\end{equation}
We numerically differentiate our time and space resolved data $\tilde{y}(\tilde{s},\tilde{t})$ and use the small deformation approximation $\tilde{s} \approx \tilde{x}$ to replace $\tilde{y}(\tilde{x},\tilde{t})$ by $\tilde{y}(\tilde{s},\tilde{t})$.
The second derivative in time and fourth derivative in space provide the left hand side of \cref{eq:momentumbalance}, which is equal to the spanwise-averaged pressure estimate ${<\tilde{p}>}_{\tilde{z}}$ multiplied by the mass ratio $M^*$ on the right hand side.
The spanwise-averaged pressure estimate ${<\tilde{p}>}_{\tilde{z}}$ is a space-time field of the $\tilde{s}$ and $\tilde{t}$ variables.
To compare different measurement time series, we take the euclidean norm in space and time over the recorded duration $\kindex{T}{r}$:
\begin{equation}
	\label{eq:spacetimeintegration}
	\| {<\tilde{p}>}_{\tilde{z}} \|_{\tilde{s},\tilde{t}} = \left( \frac{L}{\kindex{T}{r}\Uinf} \int_{\tilde{t}=0}^{\tilde{t}=\kindex{T}{r}\Uinf/L} \int_{\tilde{s}=0}^{\tilde{s}=1} <\tilde{p}>^2_{\tilde{z}}  d\tilde{s} d\tilde{t}  \right)^{1/2}.
\end{equation}

\begin{figure}
\centering
\includegraphics[]{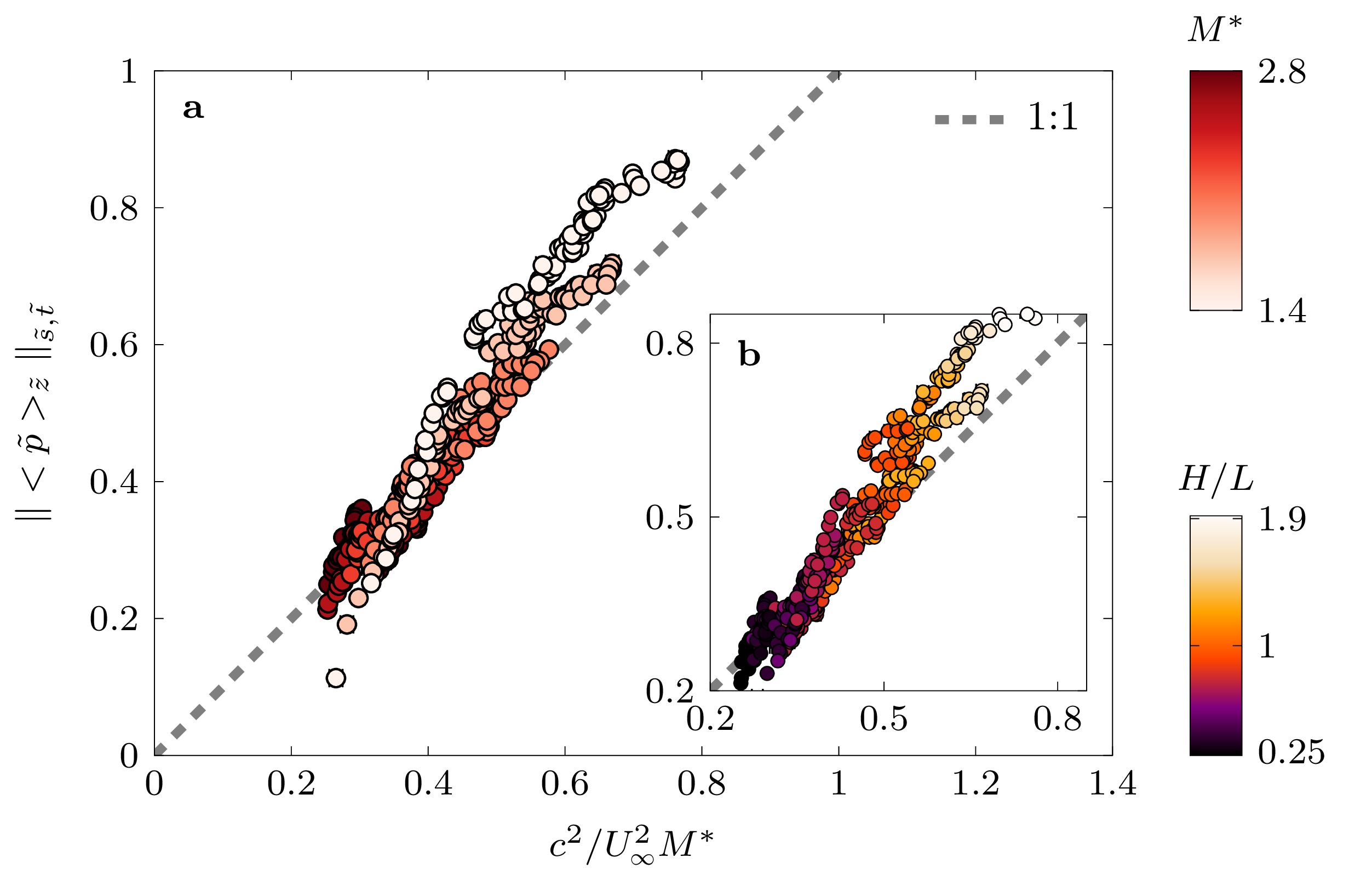}
\caption{Estimate of the overall pressure difference $\| {<\tilde{p}>}_{\tilde{z}} \|_{\tilde{s},\tilde{t}}$ as a function of the squared mass ratio corrected wave speed $c^2 / (\Uinf^2 M^*)$ coloured by (a) mass ratio $M^*$, and (b) aspect ratio $H/L$.}
\label{fig:wavespeedvspressure}
\end{figure}

We simplify \cref{eq:momentumbalance} to highlight the link between dynamic pressure and wave speed.
We follow the approach of \cite{moretti_tension_2003} that consists of (1) using an analytical approximation for the deformation and (2) neglecting the bending term of the momentum equation.
The analytical approximation for the deformation $\tilde{y}$ is a sinusoidal travelling wave modulated by a space-varying amplitude
\begin{equation}
	\label{eq:analyticaldeformation}
	\tilde{y} (\tilde{x},\tilde{t}) = \frac{A\tilde{x}}{L} \sin\left(\frac{2\pi}{\tilde{\lambda}}(\tilde{c}\tilde{t}-\tilde{x})\right), 
\end{equation}
where $A$ is the tip amplitude, $\tilde{\lambda} = \lambda/L$ is the dimensionless wavelength and $\tilde{c} = c/\Uinf$ is the dimensionless wave speed.
The simplification of a linear amplitude growth in \cref{eq:analyticaldeformation} is justified by \cite{moretti_tension_2003} citing the envelope data from \cite{watanabe_experimental_2002}.
The linear approximation does not reproduce the two dips around the necks present in the amplitude growth data presented in \cref{fig:EnvelopeAndBeamModes}c and e, but this simplification captures the overall behavior of the increasing envelope from root to tip.
The expression for the deformation of \cref{eq:analyticaldeformation} is injected in the left-hand side of the momentum balance of \cref{eq:momentumbalance}.
To simplify the expression on the left-hand side of \cref{eq:momentumbalance}, we focus on large reduced velocities corresponding to flags well in the post-critical regime (typically $U^* > 20$ and larger).
If the dimensionless time and space derivatives are the order of unity, the $1/{U^{*2}} \sim 1/400$ factor makes the bending term $\frac{1}{U^{*2}} \frac{\partial^4 \tilde{y}}{\partial \tilde{x}^4}$ negligible compared to the acceleration term $\frac{\partial^2 \tilde{y}}{\partial\tilde{t}^2}$ in \cref{eq:momentumbalance}.
The momentum balance can then be written as
\begin{equation}
\frac{\partial^2 \tilde{y}}{\partial\tilde{t}^2}  = - \frac{A\tilde{x}}{L} \left( \frac{2\pi \tilde{c}}{\tilde{\lambda}} \right)^2 \sin\left(\frac{2\pi}{\tilde{\lambda}}(\tilde{c}\tilde{t}-\tilde{x})\right) = M^* {<\tilde{p}>}_{\tilde{z}}
\end{equation}
for which we take the euclidean norm over space and the recorded duration
\begin{equation}
	\label{eq:normofomentumequation}
	\frac{A}{L} \left( \frac{2\pi \tilde{c}}{\tilde{\lambda}} \right)^2 \norm{ \tilde{x} \sin\left(\frac{2\pi}{\tilde{\lambda}}(\tilde{c}\tilde{t}-\tilde{x})\right) }_{\tilde{x},\tilde{t}} = M^* \norm{ {<\tilde{p}>}_{\tilde{z}} }_{\tilde{x},\tilde{t}}.
\end{equation}
The right hand side is our pressure estimate introduced in \cref{eq:spacetimeintegration} and the left hand side contains the properties of the wave approximation.
The norm of the $\norm{ \tilde{x} \sin\left(\frac{2\pi}{\tilde{\lambda}}(\tilde{c}\tilde{t}-\tilde{x})\right) }_{\tilde{x},\tilde{t}}$ depends on $\tilde{\lambda}$ but is overall of the order of unity.
We rewrite \cref{eq:normofomentumequation} using the dimensional values and taking the norm of the envelope times the sine as a constant factor: 
\begin{equation}
	\label{eq:wavespeedscaling}
	\frac{1}{M^*} \left( \frac{c}{\Uinf} \right)^2 \frac{A}{L} \propto \left( \frac{\lambda}{L} \right)^2 {<\tilde{p}>}_{\tilde{z}}.
\end{equation}
When we use the constant dimensionless wavelength $\lambda/L = \num{1.87}$ and neglect the tip amplitude variations (assuming $A/L\approx 0.3$), this relationship simplifies into:
\begin{equation}
\label{eq:wavespeedscalingsimplified}
c/\Uinf \propto \sqrt{M^* {<\tilde{p}>}_{\tilde{z}}}.
\end{equation}
In \cref{fig:wavespeedvspressure} we show the norm of the estimated pressure difference as a function of the square of the measured dimensionless wave speed divided by the mass ratio.
The linear variation between the two quantities supports the relationship derived in \cref{eq:wavespeedscalingsimplified}.
The data points in \cref{fig:wavespeedvspressure}a and b are coloured by mass ratio and aspect ratio, respectively.
The dynamic pressure is weakened for lower aspect ratio flags (\cref{fig:wavespeedvspressure}b) as predicted in analytical models. 
The influence of the mass ratio is less direct (\cref{fig:wavespeedvspressure}a), but when we plot the modified dimensionless wave speed  $c / (\Uinf \sqrt{M^*})$ as a function of the aspect ratio, the data from different mass ratios collapses (\cref{fig:ARTimingsA}c).
This collapse confirms that the mass corrected dimensionless wave speed directly connects the flapping response to the geometric parameters and accounts for the slow-down of the deformation wave for lower aspect ratios and mass ratios.

\section{Calculation of the circulation shed from the flag tip}\label{app:circulation}

\begin{figure}
\includegraphics[width=\linewidth]{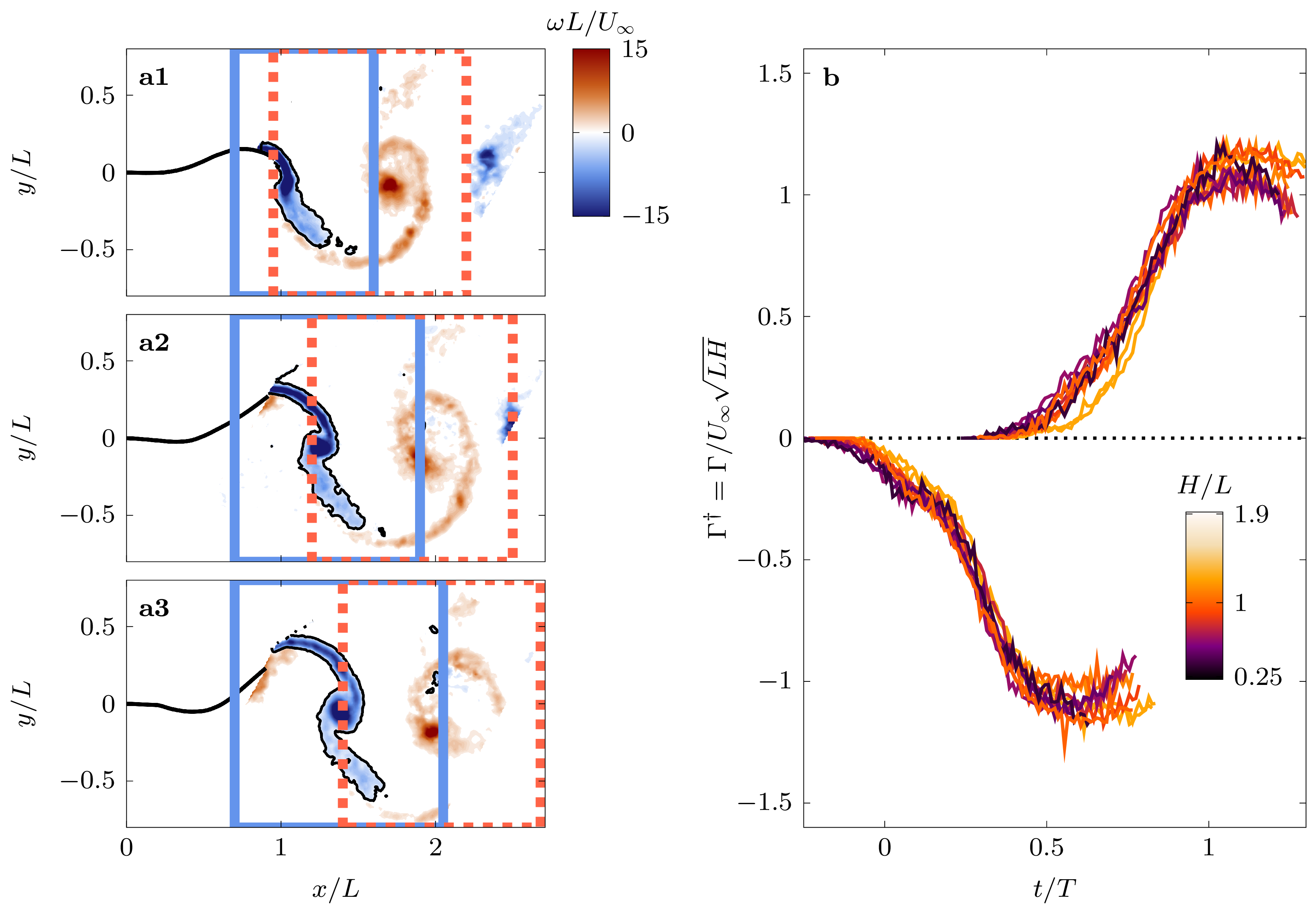}
\caption{(a1-a3) Selected instantaneous vorticity snapshots in the wake of a flapping flag.
The blue solid lines (resp. red dashed lines) delimitate the boxes inside which negative vorticity (resp. positive) is isolated.
The black contours indicate the integration domain $\kindex{S}{-}(t)$.
(b) Scaling of the evolution of the positive and negative circulation released during one flapping cycle using the square root of area $\Gamma^{\dag} = \Gamma/\Uinf \sqrt{LH}$ }
\label{fig:CirculationScalingAPpendix}
\end{figure}

The circulation data presented in \cref{subsec:wakequantities} were obtained from a manual selection of positive and negative vorticity integration domains $\kindex{S}{+}(t)$ and $\kindex{S}{-}(t)$, for individual phase instants $t$.
The positive circulation is obtained by integrating the vorticity within the manually selected time-varying integration domain $\kindex{S}{+}(t)$:
\begin{equation}
\Gamma^+ (t) = \iint_{\kindex{S}{+}(t)} \omega(x,y,t) \dd x \dd y.
\end{equation}
To obtain the integration domain $S(t)$ for each phase instant and for the positive and negative vorticity, we first manually draw a box enclosing the vorticity shed during the current stroke of the flag.
Then, we define an integration domain inside that box where the vorticity is only of the desired sign. 
In \cref{fig:CirculationScalingAPpendix}a1-a3, the blue solid line shows the box corresponding to the negative vorticity during a phase where the negative vorticity is still being released in the wake.
The red dashed line shows the box for the positive vorticity during a phase where positive vorticity is simply being convected downstream.
In both cases, the box covers the full height of the field of view.
During the phase where vorticity is released from the tip, the upstream edge of the box sits upstream of the tip, and the most downstream edge of the box is placed manually such that any vorticity of the corresponding sign released during the current stroke is included, but vorticity released during a previous stroke is excluded.
During the phase where vorticity of the corresponding sign is no longer released from the tip, the box delimits the most upstream and downstream visible patch of vorticity. 
The contour of the integration domain $\kindex{S}{-}(t)$ for the negative circulation is shown with a black line in \cref{fig:CirculationScalingAPpendix}a1-a3.
At \cref{fig:CirculationScalingAPpendix}a1, the tip has passed the $y=0$ displacement and this corresponds to a stage where negative vorticity is released at a high rate.
In \cref{fig:CirculationScalingAPpendix}a2-a3, the tip has reversed it motion direction, but there is still residual negative vorticity that is shed and that we take into account in the integration.
In \cref{fig:CirculationScalingAPpendix}a1, the previous negative vortex core, shed a cycle before, is visible at $x/L>2$ and progressively moves out of the field of view in \cref{fig:CirculationScalingAPpendix}a2 and a3.
By manually selecting the box for the integration contours, we make sure that the vorticity from the previous cycle is not taken into account in the calculation of the circulation of the current cycle.
The evolution of the positive and negative circulations, scaled with the square root of area $\sqrt{LH}$ is shown in \cref{fig:CirculationScalingAPpendix}b as a function of $t/T$.
Similarly to the scaling with $L^*$, we observe a collapse of the different curves for flags of various aspect ratio.

\section{Force data}\label{appForceData}
\begin{figure}
\includegraphics[width=\linewidth]{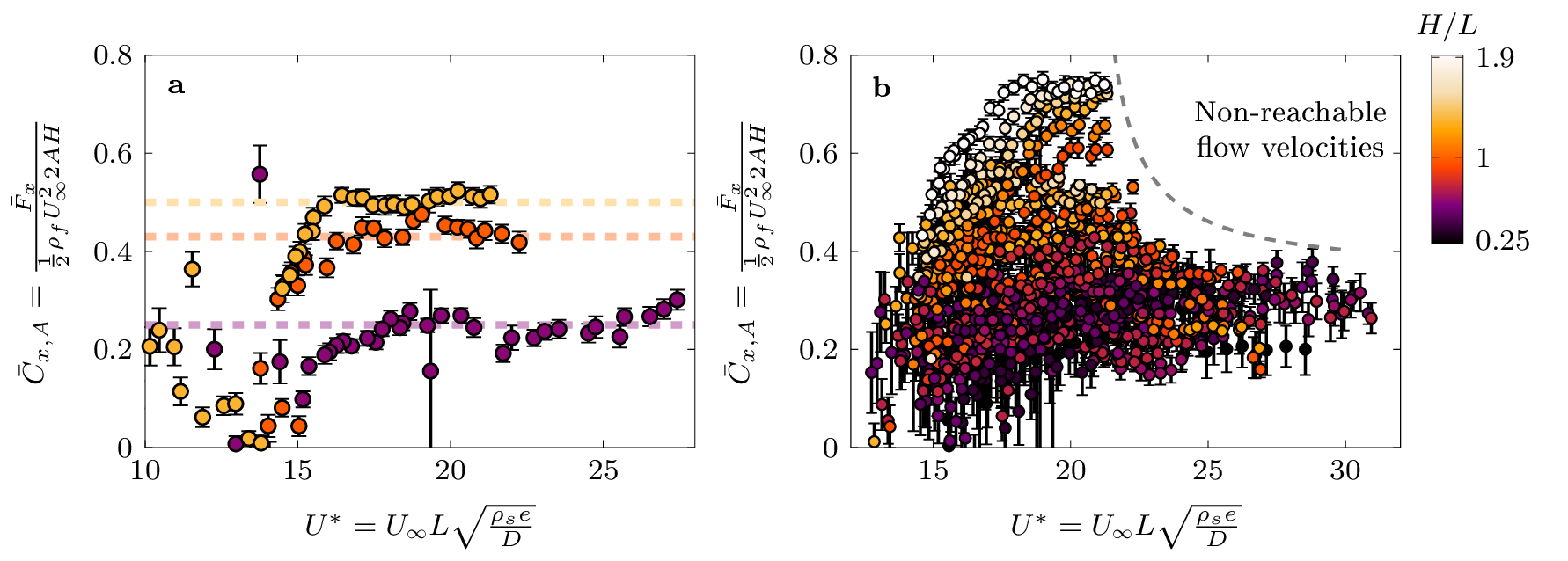}
\caption{Projected area drag coefficient for (a) the three flags of same area in \cref{fig:drag}a and (b) for all the different flags and flow velocities.}
\label{fig:DragForceDataAppendix}
\end{figure}

The choice of scaling area for the drag coefficient is traditionally the non-deformed surface $HL$.
The frontal area seen by the flow is estimated using the flapping amplitude and the height $2A\times H$.
The projected-area mean drag coefficient is defined as
\begin{equation}
	\kindex{\bar{C}}{x,A} = \frac{\kindex{\bar{F}}{x}}{\frac{1}{2} \kindex{\rho}{f} \Uinf^2 2AH} = \kindex{\bar{C}}{x} \times \frac{L}{2A},
\end{equation}
and the evolution with reduced velocity for the three examples of \cref{fig:drag}a and the full dataset are displayed in \cref{fig:DragForceDataAppendix}a and b.
Values of $\kindex{\bar{C}}{x,A}$ before the onset of flapping are heavily scattered due to the small values both for the amplitude and the mean drag.
The transition region shortly after the flapping onset is closer to the plateau because the amplitude also varies during this transition region.
The values of the projected-area mean drag coefficient at the plateau are still distinct ($\kindex{\bar{C}}{x,A} = \numlist{0.25;0.43;0.5}$).
The difference in flapping amplitude or projected area does not explain the scattering of mean drag coefficient.

\bibliographystyle{jfm}
\bibliography{bibliography}

@article{mettelsiefen.2025,
  title = {Investigating the Significance of Flatness Defects in the Origin of Hysteresis in Flag Flutter},
  author = {Mettelsiefen, Holger and Sarkar, Sunetra and Raghav, Vrishank},
	year= {2025},
  date = {2025-10-25},
  journaltitle = {Journal of Fluid Mechanics},
  shortjournal = {J. Fluid Mech.},
  volume = {1021},
  pages = {A38},
  issn = {0022-1120, 1469-7645},
  doi = {10.1017/jfm.2025.10692},
}

@article{bird_unsteady_2021,
	title = {Unsteady lifting-line theory and the influence of wake vorticity on aerodynamic loads},
	volume = {35},
	issn = {1432-2250},
	url = {https://doi.org/10.1007/s00162-021-00578-8},
	doi = {10.1007/s00162-021-00578-8},
	number = {5},
	urldate = {2025-09-25},
	journal = {Theoretical and Computational Fluid Dynamics},
	author = {Bird, Hugh J. A. and Ramesh, Kiran},
	year = {2021},
	pages = {609--631},
}

@article{bird_usefulness_2022,
	title = {Usefulness of {Inviscid} {Linear} {Unsteady} {Lifting}-{Line} {Theory} for {Viscous} {Large}-{Amplitude} {Problems}},
	volume = {60},
	issn = {0001-1452},
	url = {https://arc.aiaa.org/doi/10.2514/1.J060808},
	doi = {10.2514/1.J060808},
	number = {2},
	urldate = {2025-09-25},
	journal = {AIAA Journal},
	author = {Bird, Hugh J. A. and Ramesh, Kiran and Ōtomo, Shūji and Maria Viola, Ignazio},
	year = {2022},
	note = {Publisher: American Institute of Aeronautics and Astronautics},
	pages = {598--609},
}

@article{guermond_unified_1991,
	title = {A unified unsteady lifting-line theory},
	volume = {229},
	issn = {1469-7645, 0022-1120},
	urldate = {2025-09-25},
	journal = {Journal of Fluid Mechanics},
	author = {Guermond, Jean-Luc and Sellier, Antoine},
	month = aug,
	year = {1991},
	pages = {427--451},
}

@book{dimitriadis_unsteady_2024,
	address = {Hoboken, NJ},
	series = {Aerospace series},
	title = {Unsteady aerodynamics: potential and vortex methods},
	volume = {1},
	isbn = {978-1-119-76253-9 978-1-119-76255-3 978-1-119-76256-0},
	shorttitle = {Unsteady aerodynamics},
	language = {eng},
	publisher = {John Wiley \& Sons, Inc},
	author = {Dimitriadis, Grigorios},
	collaborator = {{EBSCOhost}},
	year = {2024},
}

@article{raynaud_event-based_2025,
	title = {Event-based reconstruction of time-resolved centreline deformation of flapping flags},
	issn = {0957-0233},
	url = {http://iopscience.iop.org/article/10.1088/1361-6501/adc1e7},
	doi = {10.1088/1361-6501/adc1e7},
	language = {en},
	urldate = {2025-03-26},
	journal = {Measurement Science and Technology},
	author = {Raynaud, Gaétan and Mulleners, Karen},
	year = {2025},
}

@article{eloy_flutter_2007,
	title = {Flutter of a rectangular plate},
	volume = {23},
	issn = {0889-9746},
	url = {https://www.sciencedirect.com/science/article/pii/S088997460700014X},
	doi = {10.1016/j.jfluidstructs.2007.02.002},
	number = {6},
	urldate = {2023-08-29},
	journal = {Journal of Fluids and Structures},
	author = {Eloy, C. and Souilliez, C. and Schouveiler, L.},
	month = aug,
	year = {2007},
	pages = {904--919},
}

@article{eloy_origin_2012,
	title = {The origin of hysteresis in the flag instability},
	volume = {691},
	issn = {1469-7645, 0022-1120},
	url = {https://www.cambridge.org/core/journals/journal-of-fluid-mechanics/article/abs/origin-of-hysteresis-in-the-flag-instability/2200F62A0E75256920925CDFAC7B487D#},
	doi = {10.1017/jfm.2011.494},
	language = {en},
	urldate = {2022-12-15},
	journal = {Journal of Fluid Mechanics},
	author = {Eloy, Christophe and Kofman, Nicolas and Schouveiler, Lionel},
	month = jan,
	year = {2012},
	note = {Publisher: Cambridge University Press},
	pages = {583--593},
}

@article{eloy_aeroelastic_2008,
	title = {Aeroelastic instability of cantilevered flexible plates in uniform flow},
	volume = {611},
	issn = {1469-7645, 0022-1120},
	url = {https://www.cambridge.org/core/journals/journal-of-fluid-mechanics/article/aeroelastic-instability-of-cantilevered-flexible-plates-in-uniform-flow/2869F573B5B1533648AFC9C6B738CF62},
	doi = {10.1017/S002211200800284X},
	language = {en},
	urldate = {2022-11-30},
	journal = {Journal of Fluid Mechanics},
	author = {Eloy, Christophe and Lagrange, Romain and Souilliez, Claire and Schouveiler, Lionel},
	month = sep,
	year = {2008},
	note = {Publisher: Cambridge University Press},
	pages = {97--106},
}

@article{virot_fluttering_2013,
	title = {Fluttering flags: {An} experimental study of fluid forces},
	volume = {43},
	issn = {0889-9746},
	shorttitle = {Fluttering flags},
	url = {https://www.sciencedirect.com/science/article/pii/S0889974613001977},
	doi = {10.1016/j.jfluidstructs.2013.09.012},
	language = {en},
	urldate = {2022-10-25},
	journal = {Journal of Fluids and Structures},
	author = {Virot, Emmanuel and Amandolese, Xavier and Hémon, Pascal},
	month = nov,
	year = {2013},
	pages = {385--401},
}

@book{raffel_particle_2007,
	address = {Berlin, Heidelberg},
	title = {Particle {Image} {Velocimetry}: {A} {Practical} {Guide}},
	copyright = {http://www.springer.com/tdm},
	isbn = {978-3-540-72307-3 978-3-540-72308-0},
	shorttitle = {Particle {Image} {Velocimetry}},
	url = {http://link.springer.com/10.1007/978-3-540-72308-0},
	language = {en},
	urldate = {2025-03-26},
	publisher = {Springer},
	author = {Raffel, Markus and Willert, Christian E. and Wereley, Steve T. and Kompenhans, Jürgen},
	year = {2007},
	doi = {10.1007/978-3-540-72308-0},
}

@article{argentina_fluid-flow-induced_2005,
	title = {Fluid-flow-induced flutter of a flag},
	volume = {102},
	url = {https://www.pnas.org/doi/abs/10.1073/pnas.0408383102},
	doi = {10.1073/pnas.0408383102},
	number = {6},
	urldate = {2024-10-15},
	journal = {Proceedings of the National Academy of Sciences},
	author = {Argentina, Médéric and Mahadevan, L.},
	month = feb,
	year = {2005},
	note = {Publisher: Proceedings of the National Academy of Sciences},
	pages = {1829--1834},
}

@article{shelley_heavy_2005,
	title = {Heavy {Flags} {Undergo} {Spontaneous} {Oscillations} in {Flowing} {Water}},
	volume = {94},
	url = {https://link.aps.org/doi/10.1103/PhysRevLett.94.094302},
	doi = {10.1103/PhysRevLett.94.094302},
	number = {9},
	urldate = {2022-11-30},
	journal = {Physical Review Letters},
	author = {Shelley, Michael and Vandenberghe, Nicolas and Zhang, Jun},
	month = mar,
	year = {2005},
	note = {Publisher: American Physical Society},
	pages = {094302},
}

@article{yu_review_2019,
	title = {A {Review} on {Fluid}-{Induced} {Flag} {Vibrations}},
	volume = {71},
	issn = {0003-6900},
	url = {https://doi.org/10.1115/1.4042446},
	doi = {10.1115/1.4042446},
	number = {010801},
	urldate = {2023-10-02},
	journal = {Applied Mechanics Reviews},
	author = {Yu, Yuelong and Liu, Yingzheng and Amandolese, Xavier},
	month = jan,
	year = {2019},
}

@article{huang_three-dimensional_2010,
	title = {Three-dimensional simulation of a flapping flag in a uniform flow},
	volume = {653},
	issn = {0022-1120, 1469-7645},
	doi = {10.1017/S0022112010000248},
	language = {en},
	urldate = {2024-01-24},
	journal = {Journal of Fluid Mechanics},
	author = {Huang, Wei-Xi and Sung, Hyung Jin},
	month = jun,
	year = {2010},
	pages = {301--336},
}

@article{thoma_schlenkernde_1939,
	title = {Das schlenkernde {Seil}},
	volume = {19},
	copyright = {Copyright © 1939 WILEY-VCH Verlag GmbH \& Co. KGaA, Weinheim},
	issn = {1521-4001},
	url = {https://onlinelibrary.wiley.com/doi/abs/10.1002/zamm.19390190508},
	doi = {10.1002/zamm.19390190508},
	language = {en},
	number = {5},
	urldate = {2024-10-11},
	journal = {ZAMM - Journal of Applied Mathematics and Mechanics / Zeitschrift für Angewandte Mathematik und Mechanik},
	author = {Thoma, D.},
	year = {1939},
	note = {\_eprint: https://onlinelibrary.wiley.com/doi/pdf/10.1002/zamm.19390190508},
	pages = {320--321},
}

@article{moretti_tension_2003,
	title = {Tension in fluttering flags},
	volume = {8},
	url = {http://www.iiav.org/ijav/content/volumes/8_2003_710551272264886/vol_4/361_fullpaper_1462641286808970.pdf},
	number = {4},
	urldate = {2024-01-24},
	journal = {International Journal of Acoustics and Vibration},
	author = {Moretti, Peter M.},
	year = {2003},
	pages = {227--230},
}

@article{connell_flapping_2007,
	title = {Flapping dynamics of a flag in a uniform stream},
	volume = {581},
	issn = {0022-1120, 1469-7645},
	url = {https://www.cambridge.org/core/product/identifier/S0022112007005307/type/journal_article},
	doi = {10.1017/S0022112007005307},
	language = {en},
	urldate = {2023-01-30},
	journal = {Journal of Fluid Mechanics},
	author = {Connell, Benjamin S. H. and Yue, Dick K. P.},
	month = jun,
	year = {2007},
	pages = {33--67},
}

@article{jankee_influence_2022,
	title = {Influence of geometrical parameters on the hysteresis of flutter onset in confined configurations},
	volume = {63},
	issn = {1432-1114},
	url = {https://doi.org/10.1007/s00348-022-03532-4},
	doi = {10.1007/s00348-022-03532-4},
	language = {en},
	number = {12},
	urldate = {2023-01-04},
	journal = {Experiments in Fluids},
	author = {Jankee, Girish K. and Ganapathisubramani, Bharathram},
	month = nov,
	year = {2022},
	pages = {183},
}

@article{gavriely_flutter_1989,
	title = {Flutter in flow-limited collapsible tubes: a mechanism for generation of wheezes},
	volume = {66},
	issn = {8750-7587},
	shorttitle = {Flutter in flow-limited collapsible tubes},
	url = {https://journals.physiology.org/doi/abs/10.1152/jappl.1989.66.5.2251},
	doi = {10.1152/jappl.1989.66.5.2251},
	number = {5},
	urldate = {2025-06-23},
	journal = {Journal of Applied Physiology},
	author = {Gavriely, N. and Shee, T. R. and Cugell, D. W. and Grotberg, J. B.},
	month = may,
	year = {1989},
	note = {Publisher: American Physiological Society},
	pages = {2251--2261},
}

@article{auregan_snoring_1995,
	title = {Snoring: linear stability analysis and in-vitro experiments},
	volume = {188},
	issn = {0022-460X},
	shorttitle = {{SNORING}},
	url = {https://www.sciencedirect.com/science/article/pii/S0022460X85705779},
	doi = {10.1006/jsvi.1995.0577},
	number = {1},
	urldate = {2024-03-05},
	journal = {Journal of Sound and Vibration},
	author = {Aurégan, Y. and Depollier, C.},
	month = nov,
	year = {1995},
	pages = {39--53},
}

@article{bornemann_leaflet_2025,
	title = {Leaflet fluttering changes laminar–turbulent transition mechanisms past bioprosthetic aortic valves},
	volume = {37},
	url = {https://pubs.aip.org/aip/pof/article/37/5/051911/3347335},
	number = {5},
	urldate = {2025-06-23},
	journal = {Physics of Fluids},
	author = {Bornemann, Karoline-Marie and Obrist, Dominik},
	year = {2025},
	note = {Publisher: AIP Publishing},
}

@book{various_encyclopaedia_1910,
	edition = {11},
	title = {Encyclopaedia {Britannica}, 11th {Edition}, "{Finland}" to "{Fleury}, {Andre}" {Volume} 10, {Slice} 4},
	volume = {10},
	isbn = {978-1-02-339164-1},
	url = {http://www.gutenberg.org/ebooks/35606},
	language = {eng},
	urldate = {2025-06-24},
	publisher = {Horace Everett Hooper},
	author = {{Various}},
	year = {1910},
}

@article{watanabe_experimental_2002,
	title = {An experimental study of paper flutter},
	volume = {16},
	issn = {0889-9746},
	url = {https://www.sciencedirect.com/science/article/pii/S0889974601904359},
	doi = {10.1006/jfls.2001.0435},
	language = {en},
	number = {4},
	urldate = {2022-12-15},
	journal = {Journal of Fluids and Structures},
	author = {Watanabe, Y. and Suzuki, S. and Sugihara, M. and Sueoka, Y.},
	month = may,
	year = {2002},
	pages = {529--542},
}

@article{huang_flutter_1995,
	title = {Flutter of {Cantilevered} {Plates} in {Axial} {Flow}},
	volume = {9},
	issn = {0889-9746},
	url = {https://www.sciencedirect.com/science/article/pii/S0889974685710079},
	doi = {10.1006/jfls.1995.1007},
	number = {2},
	urldate = {2024-10-15},
	journal = {Journal of Fluids and Structures},
	author = {Huang, L.},
	month = feb,
	year = {1995},
	pages = {127--147},
}

@article{shelley_flapping_2011,
	title = {Flapping and {Bending} {Bodies} {Interacting} with {Fluid} {Flows}},
	volume = {43},
	url = {https://doi.org/10.1146/annurev-fluid-121108-145456},
	doi = {10.1146/annurev-fluid-121108-145456},
	number = {1},
	urldate = {2023-10-30},
	journal = {Annual Review of Fluid Mechanics},
	author = {Shelley, Michael J. and Zhang, Jun},
	year = {2011},
	note = {\_eprint: https://doi.org/10.1146/annurev-fluid-121108-145456},
	pages = {449--465},
}

@article{liu_development_2012,
	title = {Development of piezoelectric microcantilever flow sensor with wind-driven energy harvesting capability},
	volume = {100},
	issn = {0003-6951},
	url = {https://doi.org/10.1063/1.4723846},
	doi = {10.1063/1.4723846},
	number = {22},
	urldate = {2025-06-24},
	journal = {Applied Physics Letters},
	author = {Liu, Huicong and Zhang, Songsong and Kathiresan, Ramprakash and Kobayashi, Takeshi and Lee, Chengkuo},
	month = may,
	year = {2012},
	pages = {223905},
}

@article{doare_piezoelectric_2011,
	title = {Piezoelectric coupling in energy-harvesting fluttering flexible plates: linear stability analysis and conversion efficiency},
	volume = {27},
	issn = {0889-9746},
	shorttitle = {Piezoelectric coupling in energy-harvesting fluttering flexible plates},
	url = {https://www.sciencedirect.com/science/article/pii/S0889974611000703},
	doi = {10.1016/j.jfluidstructs.2011.04.008},
	number = {8},
	urldate = {2025-06-24},
	journal = {Journal of Fluids and Structures},
	author = {Doaré, Olivier and Michelin, Sébastien},
	month = nov,
	year = {2011},
	pages = {1357--1375},
}

@article{rips_enhanced_2019,
	title = {Enhanced mixing at inertial microscales using flow-induced flutter},
	volume = {4},
	url = {https://link.aps.org/doi/10.1103/PhysRevFluids.4.054501},
	doi = {10.1103/PhysRevFluids.4.054501},
	number = {5},
	urldate = {2025-06-24},
	journal = {Physical Review Fluids},
	author = {Rips, Aaron and Mittal, Rajat},
	month = may,
	year = {2019},
	note = {Publisher: American Physical Society},
	pages = {054501},
}

@article{shoele_computational_2014,
	title = {Computational study of flow-induced vibration of a reed in a channel and effect on convective heat transfer},
	volume = {26},
	issn = {1070-6631},
	url = {https://doi.org/10.1063/1.4903793},
	doi = {10.1063/1.4903793},
	number = {12},
	urldate = {2025-06-24},
	journal = {Physics of Fluids},
	author = {Shoele, Kourosh and Mittal, Rajat},
	month = dec,
	year = {2014},
	pages = {127103},
}

@article{yu_energy_2016,
	title = {Energy harvesting with two parallel pinned piezoelectric membranes in fluid flow},
	volume = {65},
	issn = {0889-9746},
	url = {https://www.sciencedirect.com/science/article/pii/S0889974616301554},
	doi = {10.1016/j.jfluidstructs.2016.06.012},
	urldate = {2025-06-24},
	journal = {Journal of Fluids and Structures},
	author = {Yu, Yuelong and Liu, Yingzheng},
	month = aug,
	year = {2016},
	pages = {381--397},
}

@book{paidoussis_fluid-structure_2014,
	address = {Oxford},
	title = {Fluid-{Structure} {Interactions}},
	isbn = {978-0-12-397312-2},
	url = {https://www.sciencedirect.com/science/article/pii/B9780123973122000089},
	urldate = {2025-06-24},
	publisher = {Academic Press},
	edition = {Second Edition},
	author = {Paidoussis, Michael P.},
	month = jan,
	year = {2014},
	pages = {7-62},
	volume  = {1},
	doi = {10.1016/B978-0-12-397312-2.00008-9},
}

@article{alben_dynamics_2022,
	title = {Dynamics of flags over wide ranges of mass and bending stiffness},
	volume = {7},
	url = {https://link.aps.org/doi/10.1103/PhysRevFluids.7.013903},
	doi = {10.1103/PhysRevFluids.7.013903},
	number = {1},
	urldate = {2023-03-08},
	journal = {Physical Review Fluids},
	author = {Alben, Silas},
	month = jan,
	year = {2022},
	note = {Publisher: American Physical Society},
	pages = {013903},
}

@article{kumar_effect_2024,
	title = {The effect of aspect ratio and mass ratio on the flow-induced flutter of a thin flexible sheet},
	volume = {36},
	issn = {1070-6631},
	url = {https://doi.org/10.1063/5.0232637},
	doi = {10.1063/5.0232637},
	number = {11},
	urldate = {2025-04-22},
	journal = {Physics of Fluids},
	author = {Kumar, Dhiraj and Poddar, Kamal and Kumar, Sanjay},
	month = nov,
	year = {2024},
	pages = {114109},
}

@article{taneda_waving_1968,
	title = {Waving {Motions} of {Flags}},
	volume = {24},
	issn = {0031-9015},
	url = {https://journals.jps.jp/doi/10.1143/JPSJ.24.392},
	doi = {10.1143/JPSJ.24.392},
	number = {2},
	urldate = {2022-10-12},
	journal = {Journal of the Physical Society of Japan},
	author = {Taneda, Sadatoshi},
	month = feb,
	year = {1968},
	note = {Publisher: The Physical Society of Japan},
	pages = {392--401},
}

@article{kumar_dynamics_2021,
	title = {The dynamics of flow-induced flutter of a thin flexible sheet},
	volume = {33},
	issn = {1070-6631},
	url = {https://doi.org/10.1063/5.0042617},
	doi = {10.1063/5.0042617},
	number = {3},
	urldate = {2025-04-22},
	journal = {Physics of Fluids},
	author = {Kumar, Dhiraj and Arekar, Ashwini N. and Poddar, Kamal},
	month = mar,
	year = {2021},
	pages = {034131},
}

@article{caverly_invariant_2025,
	title = {Invariant scaling of impulsively started polygonal disks},
	volume = {1010},
	url = {https://www.cambridge.org/core/journals/journal-of-fluid-mechanics/article/invariant-scaling-of-impulsively-started-polygonal-disks/58A2C9F23C6A664CA3054737DA1EE7B7},
	urldate = {2025-09-16},
	journal = {Journal of Fluid Mechanics},
	author = {Caverly, Dylan and Nedić, Jovan},
	year = {2025},
	note = {Publisher: Cambridge University Press},
	pages = {R1},
}

@article{fernando_reynolds-number_2016,
	title = {Reynolds-number scaling of vortex pinch-off on low-aspect-ratio propulsors},
	volume = {799},
	issn = {0022-1120, 1469-7645},
	url = {https://www.cambridge.org/core/journals/journal-of-fluid-mechanics/article/reynoldsnumber-scaling-of-vortex-pinchoff-on-lowaspectratio-propulsors/679B46209BD7985A19E7BD9027CAEC0E},
	doi = {10.1017/jfm.2016.396},
	language = {en},
	urldate = {2025-09-16},
	journal = {Journal of Fluid Mechanics},
	author = {Fernando, John N. and Rival, David E.},
	month = jul,
	year = {2016},
	pages = {R3},
}

@article{nedic_drag_2013,
	title = {Drag and near wake characteristics of flat plates normal to the flow with fractal edge geometries},
	volume = {45},
	issn = {1873-7005},
	url = {https://dx.doi.org/10.1088/0169-5983/45/6/061406},
	doi = {10.1088/0169-5983/45/6/061406},
	language = {en},
	number = {6},
	urldate = {2025-09-16},
	journal = {Fluid Dynamics Research},
	author = {Nedić, J and Ganapathisubramani, B and Vassilicos, J C},
	month = aug,
	year = {2013},
	note = {Publisher: IOP Publishing},
	pages = {061406},
}

@article{lemaitre_instability_2005,
	series = {Fluid-{Plate} {Interactions}},
	title = {Instability of a long ribbon hanging in axial air flow},
	volume = {20},
	issn = {0889-9746},
	url = {https://www.sciencedirect.com/science/article/pii/S0889974605000733},
	doi = {10.1016/j.jfluidstructs.2005.04.009},
	number = {7},
	urldate = {2025-09-16},
	journal = {Journal of Fluids and Structures},
	author = {Lemaitre, C. and Hémon, P. and de Langre, E.},
	month = oct,
	year = {2005},
	pages = {913--925},
}

@article{michelin_vortex_2008,
	title = {Vortex shedding model of a flapping flag},
	volume = {617},
	issn = {1469-7645, 0022-1120},
	url = {https://www.cambridge.org/core/journals/journal-of-fluid-mechanics/article/vortex-shedding-model-of-a-flapping-flag/36724CF933F0EC0FAE1C12207384D553},
	doi = {10.1017/S0022112008004321},
	language = {en},
	urldate = {2022-11-30},
	journal = {Journal of Fluid Mechanics},
	author = {Michelin, Sébastien and Smith, Stefan G. Llewellyn and Glover, Beverley J.},
	month = dec,
	year = {2008},
	note = {Publisher: Cambridge University Press},
	pages = {1--10},
}

@article{muller_fish_2003,
	title = {Fish 'n {Flag}},
	volume = {302},
	url = {https://www.science.org/doi/10.1126/science.1092367},
	doi = {10.1126/science.1092367},
	number = {5650},
	urldate = {2025-09-23},
	journal = {Science},
	author = {Müller, Ulrike K.},
	month = nov,
	year = {2003},
	note = {Publisher: American Association for the Advancement of Science},
	pages = {1511--1512},
}

@article{tang_instability_2007,
	title = {On the instability and the post-critical behaviour of two-dimensional cantilevered flexible plates in axial flow},
	volume = {305},
	issn = {0022-460X},
	url = {https://www.sciencedirect.com/science/article/pii/S0022460X07002295},
	doi = {10.1016/j.jsv.2007.03.042},
	number = {1},
	urldate = {2025-09-23},
	journal = {Journal of Sound and Vibration},
	author = {Tang, Liaosha and Païdoussis, Michael P.},
	month = aug,
	year = {2007},
	pages = {97--115},
}

@article{green_unsteady_2011,
	title = {The unsteady three-dimensional wake produced by a trapezoidal pitching panel},
	volume = {685},
	issn = {1469-7645, 0022-1120},
	url = {https://www.cambridge.org/core/journals/journal-of-fluid-mechanics/article/unsteady-threedimensional-wake-produced-by-a-trapezoidal-pitching-panel/527CA5A9CFC7DE5A6171493354738D7D},
	doi = {10.1017/jfm.2011.286},
	language = {en},
	urldate = {2026-01-30},
	journal = {Journal of Fluid Mechanics},
	author = {Green, Melissa A. and Rowley, Clarence W. and Smits, Alexander J.},
	month = oct,
	year = {2011},
	pages = {117--145},
}

@article{vogel2009leaves,
  title={Leaves in the lowest and highest winds: temperature, force and shape},
  author={Vogel, Steven},
  journal={New Phytologist},
  volume={183},
  number={1},
  pages={13--26},
  year={2009},
  publisher={Wiley Online Library}
}

@article{fernando2016vortex,
  title={On vortex evolution in the wake of axisymmetric and non-axisymmetric low-aspect-ratio accelerating plates},
  author={Fernando, John N and Rival, David E},
  journal={Physics of Fluids},
  volume={28},
  number={1},
  year={2016},
  publisher={AIP Publishing}
}

@article{hiroaki2024flutter,
  title={Flutter generated on a sheet in 3D flow: influence of aspect ratio of sheets on post-critical and hysteretic behavior},
  author={Hiroaki, Keiichi and Watanabe, Masahiro},
  journal={Nonlinear Dynamics},
  volume={112},
  number={18},
  pages={15757--15770},
  year={2024},
  publisher={Springer}
}

@article{chun2010flutter,
  title={Flutter of finite-span flexible plates in uniform flow},
  author={Chun-Yu, Bao and Chao, Tang and Xie-Zhen, Yin and Xi-Yun, Lu},
  journal={Chinese Physics Letters},
  volume={27},
  number={6},
  pages={064601},
  year={2010}
}

@article{higuchi1996three,
  title={Three-dimensional wake formations behind a family of regular polygonal plates},
  author={Higuchi, H and Anderson, RW and Zhang, J},
  journal={AIAA journal},
  volume={34},
  number={6},
  pages={1138--1145},
  year={1996}
}

\end{document}